\documentclass[12pt]{article}
\usepackage[dvips]{graphicx}
\usepackage{amsfonts,amsbsy,amssymb}
\begin{document}    
\setlength{\normalbaselineskip}{20pt plus 0.2pt minus 01.pt}
\baselineskip=\normalbaselineskip
\ifx\TwoupWrites\UnDeFiNeD\else\target{\magstepminus1}{11.3in}{8.27in}
	\source{\magstep0}{7.5in}{11.69in}\fi
\newfont{\fourteencp}{cmcsc10 scaled\magstep2}
\newfont{\titlefont}{cmbx10 scaled\magstep2}
\newfont{\authorfont}{cmcsc10 scaled\magstep1}
\newfont{\fourteenmib}{cmmib10 scaled\magstep2}
	\skewchar\fourteenmib='177
\newfont{\elevenmib}{cmmib10 scaled\magstephalf}
	\skewchar\elevenmib='177
\newif\ifpUbblock  \pUbblocktrue
\newcommand\nopubblock{\pUbblockfalse}
\newcommand\topspace{\hrule height 0pt depth 0pt \vskip}
\newcommand\pUbblock{\begingroup \tabskip=\hsize minus \hsize
	\baselineskip=1.5\ht\strutbox \topspace-2\baselineskip
	\halign to\hsize{\strut ##\hfil\tabskip=0pt\crcr
	\the\Pubnum\crcr\the\date\crcr}\endgroup}
\renewcommand\titlepage{\ifx\TwoupWrites\UnDeFiNeD\null\vspace{-1.7cm}\fi
	\YITPmark\vskip0.6cm
	\ifpUbblock\pUbblock \else\hrule height 0pt \relax \fi}
\newtoks\date
\newtoks\Pubnum
\newtoks\pubnum
\Pubnum={YITP-98-66}
\date={\today}
\newcommand{\frontpageskip}{\vspace{12pt plus .5fil minus 2pt}}
\renewcommand{\title}[1]{\frontpageskip
	\begin{center}{\titlefont #1}\end{center}\par}
\renewcommand{\author}[1]{\frontpageskip\par\begin{center}
	{\authorfont #1}\end{center}
	\nobreak
	}
\newcommand{\andauthor}{\frontpageskip\centerline{and}\author}
\newcommand{\authors}{\frontpageskip\noindent}
\newcommand{\address}[1]{\par\begin{center}{\sl #1}\end{center}\par}
\newcommand{\andaddress}{\par\centerline{\sl and}\address}
\renewcommand{\thanks}[1]{\footnote{#1}}
\renewcommand{\abstract}{\par\frontpageskip\centerline{\fourteencp Abstract}
	\vspace{8pt plus 3pt minus 3pt}}
\newcommand\YITP{\address{
	       Yukawa Institute for Theoretical Physics\\Kyoto
	       University, Kyoto 606-8502, Japan\\}}
\thispagestyle{empty}
%
\newcommand{\del}{\partial}
\newcommand{\m}{\mathbf}
\newcommand{\x}{{\mathbf{x}}}
\newcommand{\y}{{\mathbf{y}}}
\newcommand{\z}{{\mathbf{z}}}
\newcommand{\n}{{\mathbf{n}}}
\newcommand{\bi}{{\mathbf{i}}}
\newcommand{\bj}{{\mathbf{j}}}
\newcommand{\bI}{{\mathbf{I}}}
\newcommand{\M}{\mathbb}
\newcommand{\B}{\boldsymbol}
\newcommand{\IG}{\includegraphics}
\newcommand{\BF}{\begin{figure}\begin{center}}
\newcommand{\EF}{\end{center}\end{figure}}
\newcommand{\BE}{\begin{equation}}
\newcommand{\EE}{\end{equation}}
\newcommand{\U}{\underline}
\newcommand{\ti}{\textit}

\newcommand{\f}{\frac}
\newcommand{\T}{\tilde}
\newcommand{\N}{\nonumber}
\newcommand{\bb}{\bibitem}

\title{Computation of eigenmodes \\on a compact hyperbolic 3-space 
}

\author{Kaiki Taro INOUE
}
\YITP

\abstract{Measurements of cosmic microwave background (CMB)
 anisotropy are ideal experiments 
for discovering the non-trivial global topology of the universe. 
To evaluate the CMB anisotropy in multiply-connected compact
cosmological models, one needs to compute the eigenmodes of the
Laplace-Beltrami operator. 
Using the direct boundary element method, we numerically obtain 
the low-lying eigenmodes on a compact 
hyperbolic 3-space called the Thurston manifold which is the second
smallest in the known compact hyperbolic 3-manifolds. The 
computed eigenmodes are 
expanded in terms of eigenmodes on the unit three-dimensional 
pseudosphere. We numerically find that the expansion coefficients 
behave as Gaussian pseudo-random numbers for low-lying eigenmodes.
The observed gaussianity in the CMB fluctuations can partially  
be attributed to the Gaussian pseudo-randomness of the expansion
coefficients assuming that the Gaussian pseudo-randomness 
is the universal property of the compact hyperbolic spaces.}
\newpage
\section{Introduction} 
\indent

In recent years, there has been a great interest in properties of CMB
anisotropy in multiply-connected cosmological 
models \cite{Ste,deO2,Horn,Flat}. 
Most of these studies deal with flat models or non-compact 
hyperbolic models for
which the eigenmodes are known explicitly. Since no closed
analytic expression of eigenmodes is known for compact
hyperbolic (CH) models, so far, analysis of the CMB anisotropy in CH models
has been considered to be quite difficult although they have 
interesting properties which are strikingly different from that of 
multiply-connected flat models.    
For instance, in low $\Omega_o$
adiabatic models, the large-angular fluctuations can be produced at
periods after the last scattering as the 
curvature perturbations decay in the curvature dominant era. Therefore, the 
argument of the suppression of the large-angular fluctuations due
to the ''mode-cutoff'' in the multiply-connected flat models cannot
simply be applicable to the multiply-connected hyperbolic models. 
\\
\indent
Because the effect of the multiply-connectedness becomes
significant as the volume of the space becomes small, it is very
important to study whether the''small'' universe scenario 
is plausible. For instance, the Weeks manifold and the 
Thurston manifold have volume$\sim$$R^3$ where $R$
denotes the curvature radius, and they are the smallest and the second 
smallest compact hyperbolic manifolds, respectively. For a technical 
reason, we study the properties of eigenmodes on the Thurston
manifold. Since each type(scalar, vector and tensor) of perturbations 
evolves independently on
the locally homogeneous and isotropic FRW background space, in
$k$-space,  
the perturbations are given by the eigenmodes and the 
time evolution of the perturbations on the locally 
homogeneous and isotropic FRW space.  The periodic
boundary conditions on the eigenmodes drastically change 
the nature of the CMB fluctuations on the topological identification
scale while on smaller scale they asymptotically converge to that
in the standard locally and globally homogeneous and isotropic 
FRW background space. Thus computation of eigenmodes are very
important in understanding the properties of CMB fluctuations
\footnote{It should be noted that the computation of 
eigenmodes is also essential in the 
framework of spectral geometry \cite{Seriu}.}.
\\
\indent 
Computing the eigenmodes of the
Laplace-Beltrami operator in CH spaces (manifolds) 
is equivalent to solving the Helmholtz 
equation with appropriate periodic boundary conditions in the 
universal covering space.
A number of numerical methods have been used for solving the 
Helmholtz equation such as the finite element methods and the 
finite difference methods \cite{Aurich,Huntebrinker,Cornish}. 
\\
\indent
A numerical method called the ''direct boundary element method'' 
(DBEM) has been  used by Aurich and Steiner for finding out
the eigenmodes of the Laplace-Beltrami operator in a 
two-dimensional compact multiply-connected
space for studying the statistical properties of the eigenmodes in
highly-excited states or 
equivalently the semi-classical wavefunctions \cite{Aur1}. 
We find that this pioneering
work for ''quantum chaology'' , the study of 
the imprints of classical chaos in the quantum
mechanical counterparts is very useful for the study 
of the CMB anisotropy in CH cosmological models as well.      
The advantage of the DBEM is that it reduces the dimensionality of the
problem by one which leads to economy in the numerical task.  
Since one needs to discretize only the boundary, generation of
meshes is much easier than the other methods. Furthermore, as we shall see 
later, the DBEM is suitable for expanding the eigenfunctions that are
continued onto the whole Poincar$\acute{\textrm{e}}$ ball by the periodic
boundary conditions in terms of
eigenfunctions on the pseudosphere\footnote{A set of the continued
eigenfunctions is a subset of all eigenfunctions on the universal
covering space.}. In order to compute a set of expansion
coefficients, one needs to compute the values of the corresponding
eigenfunction on a hyperbolic sphere with appropriate
radius. If one does not care about the normalization of the
eigenfunctions, unlike the FEM, the computation of the eigenfunctions 
on the whole fundamental domain is not 
necessary. In the DBEM,  computation of the eigenfunction at an arbitrary point 
either inside or outside of the fundamental domain 
can be done by using the values of the eigenfunction and the 
normal derivatives on the boundary. 
\\
\indent
As the classical dynamical systems in CH spaces are strongly
chaotic, one can naturally assume that the imprint of the 
classical chaos is hidden in the 
corresponding quantum systems in some way. 
It has been found that the expansion coefficients with a 
certain basis behave as if they 
are random Gaussian numbers in some 
classically chaotic systems (\cite{Aur1,Haake}), which is consistent
with the prediction of random-matrix theory(\cite{Bro}).
Since the CMB temperature fluctuations in CH spaces are written 
in terms of expansion
coefficients and the eigenfunctions on the universal covering space
plus initial fluctuations, if the random behavior of the expansion
coefficients is confirmed, 
the origin of the random gaussianity
in the CMB temperature fluctuations can be partially explained
in terms of the geometric property of the universe. 
\\
\indent
In this paper, we introduce the DBEM for solving the Helmholtz equation.
Then we apply the DBEM for computing the low-lying eigenmodes of the
Laplace-Beltrami operator in the Thurston manifold.
The computed eigenfunctions are naturally continued onto the whole 
Poincar$\acute{\textrm{e}}$ 
ball because of the periodic boundary conditions and are expanded in 
terms of eigenmodes on the simply-connected pseudosphere. 
Statistical properties 
of the expansion coefficients are examined, since they are key factors 
in understanding of the CMB anisotropy in CH models. 
\section{The direct boundary element method (DBEM)}
\indent

The boundary element methods (BEM) use free Green's function as
the weighted function, and the Helmholtz equation is
rewritten as an integral equation defined on the
boundary using Green's theorem. Discretization of the 
boundary integral equation yields a
system of linear equations. Since one needs the discretiztion on only
the boundary, BEM reduces the dimensionality of the
problem by one which leads to economy in the numerical task.   
To locate an eigenvalue, the DBEM \footnote{The DBEM uses only
boundary points in evaluating the integrand in 
Eq.(\ref{eq:re2}). The indirect
methods use internal points in evaluating the integrand in 
Eq.(\ref{eq:re2}) as well as the boundary points.} requires
one to compute many determinants of the corresponding boundary
matrices which are dependent on the wavenumber $k$. 
\\
\indent
Firstly, let us consider the Helmholtz equation with certain boundary 
conditions,
\begin{equation}
(\nabla^2+k^2)u(\x)=0,\label{eq:helmholtz}
\end{equation}
which is defined on a bounded
M-dimensional connected and simply-connected domain 
$\Omega$ which is a subspace of a 
M-dimensional Riemannian manifold ${\cal M}$ and the boundary 
$\del\Omega$ is piecewise smooth. $\nabla^2\equiv\nabla^i \nabla_i,~
(i=1,2,\cdot\cdot\cdot,M)$, and $\nabla_i$ is the covariant derivative 
operator defined on $\cal M$. A function $u$ in Sobolev space 
$H^2(\Omega)$ is the
solution of the Helmholtz equation if and only if 
\begin{equation}
{\cal R}[u(\x),v(\x)]\equiv \Bigl\langle(\nabla^2+k^2)\
u(\x),v(\x)\Bigr\rangle=0,\label{eq:residue}
\end{equation}
where $v$ is an arbitrary function in Sobolev space $H^1(\Omega)$ called 
\textit{weighted function} and $\langle\,\rangle$ is defined as 
\begin{equation}
\langle a,b \rangle\equiv \int_{\Omega}  ab\sqrt{g}\, dV. 
\end{equation}    
Next, we put $u(\x)$ into the form
\begin{equation}
u=\sum_{j=1}^M u_j \phi_j,
\end{equation}
where $\phi_j$'s are linearly independent
square-integrable functions.  Numerical
methods such as the finite element methods try 
to minimize the residue function $\cal R$ for a fixed
weighted function $v(\x)$ by changing the coefficients $u_j$. 
In these methods, one must 
resort to the variational principle to find the $u_j$'s which 
minimize ${\cal R}$.
\\
\indent
Now we formulate the DBEM
which is a version of BEMs. Here we search $u(\x)$'s for the space 
$C^1(\bar{\Omega})\cap C^2(\Omega)\cap L^2(\Omega)$. 
First, we slightly 
modify Eq.(\ref{eq:residue}) using the Green's theorem
\begin{equation}
\int_\Omega (\nabla^2 u) v \sqrt{g} dV
-\int_\Omega (\nabla^2 v) u \sqrt{g} dV
=\int_{\del\Omega} (\nabla_i u) v \sqrt{g} dS^i
-\int_{\del\Omega} (\nabla_i v) u \sqrt{g} dS^i,
\label{eq:re2}
\end{equation}
where $g \equiv \mathrm{det}\{ g_{ij} \}$ and $dV\equiv dx_1 \ldots dx_M$;
the surface element $dS^i$ is given by
\begin{eqnarray}
dS_i &\equiv& \f{1}{M !} \epsilon_{i j_1 \cdot \cdot \cdot j_M} 
dS^{j_1 \cdot \cdot \cdot j_M},
\nonumber
\\
\nonumber
\\
dS^{j_1 \ldots j_M}&\equiv&
\left|
\begin{array}{@{\,}cccc@{\,}}
dx^{(1)j_1} & dx^{(2)j_1} & \ldots & dx^{(M)j_1} \\
dx^{(1)j_2} & dx^{(2)j_2} & \ldots & dx^{(M)j_2} \\
\vdots      &   \vdots    & \ddots & \vdots      \\
dx^{(1)j_M} & dx^{(2)j_M} & \ldots & dx^{(M)j_M}
\end{array}
\right|,
\end{eqnarray}
\\
where $\epsilon_{j_1 \cdot \cdot \cdot j_{M+1}}$ denotes the M$
\!+1$-dimensional Levi-Civita tensor. 
Then Eq.(\ref{eq:residue}) becomes
\begin{equation}
\int_\Omega (\nabla^2 v+k^2 v)u  \sqrt{g}\, dV
+\int_{\del\Omega} (\nabla_i u) v \sqrt{g}\, dS^i
-\int_{\del\Omega} (\nabla_i v) u \sqrt{g}\, dS^i=0.
\label{eq:IN}
\end{equation} 
As the weighted function v, we choose the fundamental solution
$G_E(\x,\y)$ which satisfies  
\begin{equation}
(\nabla^2+E)G_E(\x,\y)=\delta_D(\x-\y),
\label{eq:DE}
\end{equation}
where $E\equiv k^2$, and $\delta_D(\x-\y)$ 
is Dirac's delta function.
$G_E(\x,\y)$ is also known as the free Green's function 
whose boundary 
condition is given as 
\BE
\lim _{d(\x,\y) \rightarrow \infty} G_E(\x,\y) =0, 
\EE
where $d(\x,\y)$ is the geodesic distance between $\x$ and $\y$.
 Let $\y$ be an internal point of $\Omega$. Then
we obtain from Eq.(\ref{eq:IN}) and Eq.(\ref{eq:DE}),
\begin{equation}
u(\y)
+\int_{\del\Omega} G_E(\x,\y) \nabla_i u \,\sqrt{g}\, dS^i
-\int_{\del\Omega} (\nabla_i G_E(\x,\y)) u \,\sqrt{g}\, dS^i=0.\label{eq:inter}
\end{equation}
Thus the values of eigenfunctions at internal points can be 
computed using only the boundary integral. If $\y \in
\del\Omega$, we have to evaluate the limit of the
boundary integral terms as  $G_E(\x,\y)$ becomes divergent at $\x=\y$
(see appendix A). The boundary integral
equation is finally written as 
\begin{equation}
\f{1}{2}u(\y)
+\int_{\del\Omega} G_E(\x,\y) \nabla_i u \,\sqrt{g}\, dS^i
-\int_{\del\Omega} (\nabla_i G_E(\x,\y)) u \,\sqrt{g}\, dS^i=0,
\label{eq:bem0}
\end{equation}
or in another form, 
\begin{equation}
\f{1}{2}u(\y)
+\int_{\del\Omega} G_E(\x,\y) \f{\del u}{\del x^i} n^i \,\sqrt{g}\, dS
-\int_{\del\Omega} \f{\del G_E(\x,\y)}{\del x^i}n^i\, u \,\sqrt{g}\, dS=0,
\label{eq:bem}
\end{equation} 
where $n^i\equiv dS^i/dS$ and $dS\equiv \sqrt{dS^i\,dS_i}$. Note that
we assumed that the boundary surface at $\y$ is sufficiently smooth.
If the boundary is not smooth, one must calculate the internal solid
angle at $\y$ (see appendix A). Another approach is to rewrite 
Eq.(\ref{eq:inter}) in a regularized form \cite{Tan}. We see from 
Eq.(\ref{eq:bem0}) or Eq.(\ref{eq:bem}) that the approximated
solutions can be obtained without resorting to the variational principle.
Since it is virtually impossible to solve Eq.(\ref{eq:bem})
analytically, we discretize it using boundary elements. Let
the number of the elements be N. We approximate $u$  by some
low-order polynomials (shape function) on each element as
$u=c_1+c_2\, \eta+c_3\, \xi$ where $\eta$ and $\xi$
denote the coordinates on the corresponding standard
element \footnote{It can be proved that the approximated 
polynomial solutions converge to $u(\x)$ as the number of boundary 
elements becomes large \cite{Tabata,Johnson}.}.
   
Then we have the following equation:
\begin{equation}
[H]\{u\}=[G]\{q\},~~~~q\equiv \f{\del u}{\del n},\label{eq:BEM}
\end{equation}
where \{u\} and \{q\} are N-dimensional vectors which consist of 
the boundary values of 
an eigenfunction and its normal derivatives, respectively. 
[H] and [G] are 
N$\times$N-\,dimensional coefficient matrices which are
obtained from integration of 
the fundamental solution $G_E(\x,\y)$ and its normal derivatives multiplied 
by $u_i$ and $q_i$, respectively. The explicit form of [H] and [G] 
for constant elements are given in section 4. Note that the elements in 
[H] and [G] include $k$ implicitly.
Because Eq.(\ref{eq:BEM}) includes both $u$ and $q$, the boundary 
element method can naturally incorporate the periodic boundary conditions: 
\begin{equation}
u(\x)=u(g_i(\x)),~~~q(\x)=-q(g_i(\x)),~~\textrm{on}~~ \del\Omega,
\label{eq:po}
\end{equation}
where $g_i$'s are the face-to-face identification maps 
defined on the boundary(see appendix B). 
The boundary
conditions constrain the number of unknown constants to N. 
Application of the boundary condition (\ref{eq:po}) to Eq.(\ref{eq:BEM})
and permutation of the columns of the 
components yields
\begin{equation}
[A]\{x\}=0, \label{eq:Ax}
\end{equation}
where  N$\times$N-\,dimensional matrix $A$ is constructed from
$G_{ij}$ and $H_{ij}$ and N-\,dimensional vector 
$x$ is constructed from $u_i$'s and $q_i$'s. For the presence of 
the non-trivial 
solution, the following relation must hold,
\begin{equation}
\mathrm{det}[A]=0. \label{eq:dA}
\end{equation}
Thus the eigenvalues of the Laplace-Beltrami operator acting on the
space $C^1(\bar{\Omega})\cap C^2(\Omega)\cap L^2(\Omega)$ are obtained by
searching for $k$'s which satisfy Eq.(\ref{eq:dA}). 
\pagebreak
\
\section{Computation of low-lying eigenmodes}
\indent

In this section, we apply the DBEM for computing
the low-lying eigenmodes on the Thurston manifold $Q_2$.    
We have chosen $Q_2$ for a technical reason that the 
fundamental domain of $Q_2$ is much simpler than that of
the Weeks manifold $Q_1$(generation of meshes is much simpler). See appendix B
for understanding the basic aspects of three-dimensional 
hyperbolic geometry. 
\\
\indent
The Helmholtz equation in the Poincar$\acute{\textrm{e}}$ coordinates 
is written as
\begin{equation}
\f{1}{4}\biggl (1-|\x|^2\biggr )^2 \Biggl
[\Delta_E+ \f{2}{1-|\x|^2}~~ \x \cdot \nabla_E  \Biggr ] u + k^2 u=0,
\label {eq:Hel}  
\end{equation}
where $\Delta_E$ and $\nabla_E$ are the Laplacian and the gradient 
on the corresponding three-dimensional Euclidean space, respectively.
Note that we have set the curvature radius $R=1$ without loss of 
generality. By using the DBEM, the Helmholtz equation (\ref{eq:Hel}) 
is converted to an integral representation on the boundary. 
Here Eq.(\ref{eq:bem})  can
be written in terms of Euclidean quantities as
\begin{equation}
\f{1}{2}u(\y)
+\int_{\del\Omega} G_k(\x,\y) \f{\del u}{\del x^i}\, n_E^i\,
  dS
-\int_{\del\Omega} \f{\del G_k(\x,\y)}{\del x^i}\,u \,n_E^i\,
 dS=0,
\label{eq:bem2}
\end{equation}
where $dS=2(1-|\x|^2)^{-1}\, dS_E$.
The fundamental solution is given as \cite{Els,Tom}
\begin{equation}
G_k\,(\x,\y)=-\f{1}{4 \pi}\,\,
\f{\Bigl( \sigma+\sqrt{\sigma^2-1}\Bigr)^{-s}}{\sqrt{\sigma^2-1}},~~~
-\f{\pi}{2} < \mathrm{arg}\,\, s \leq \f{\pi}{2},
\end{equation}  
where $s=\sqrt{1-k^2}$ and ~$\sigma=\cosh d(\x,\y)$. 
Then Eq.(\ref{eq:bem2}) is discretized on the boundary elements
$\Gamma_J$ as
\begin{equation}
\f{1}{2}u(\x_I)
+\sum_{J=1}^N \Biggl[ \int_{\Gamma_J} G_k(\x_I,\y_J) 
\f{\del u(\y_J)}{\del n}\, dS
-\int_{\Gamma_J} \f{\del G_k(\x_I,\y_J)}{\del n}u(\y_J)\, dS\Biggr]=0,
\label{eq:bem2d}
\end{equation}
where N denotes the number of the boundary elements.  An example of 
$N\!=\!1168$ elements on
the boundary of the fundamental
domain in the Poincar$\acute{\textrm{e}}$ coordinates is shown in
figure \ref{fig:1168BE}. These elements are firstly generated in Klein
coordinates in which the mesh-generation is convenient.  
\BF
\includegraphics[clip]{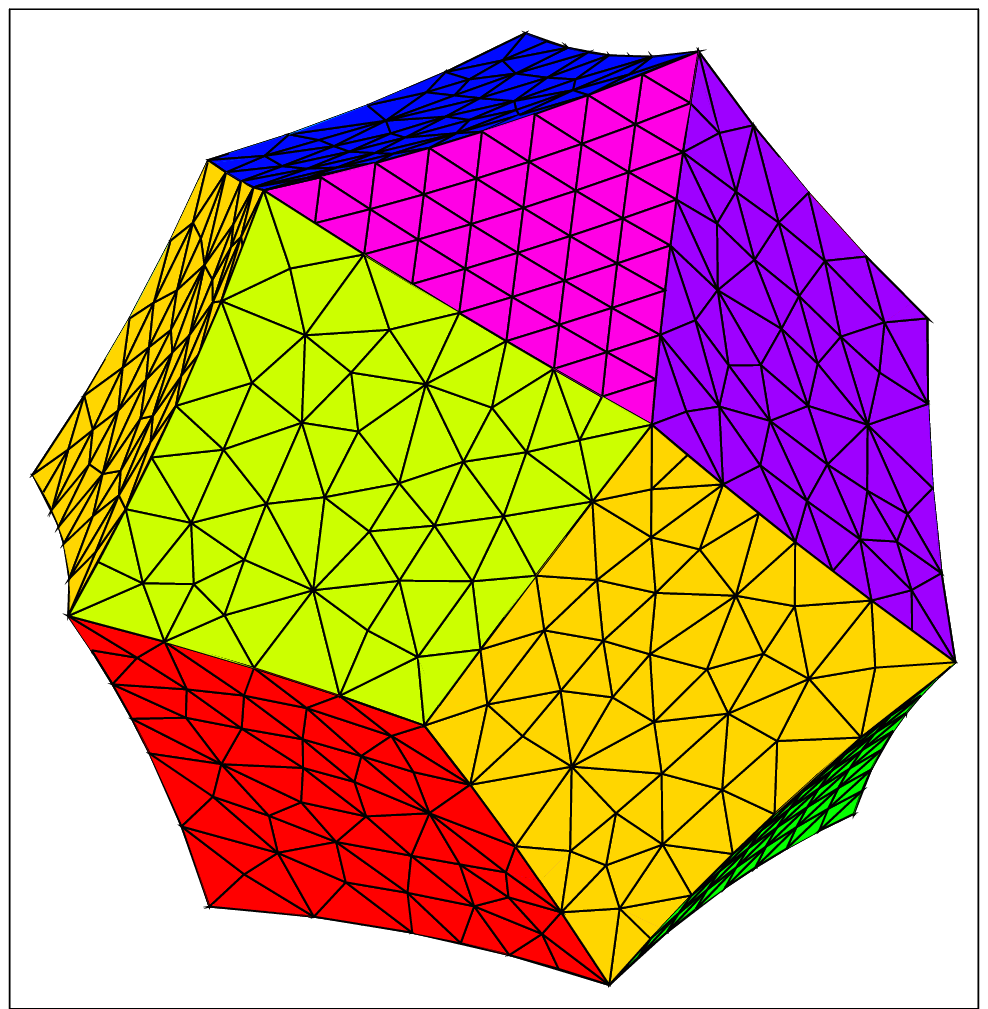}
\caption{1168 boundary elements}
\label{fig:1168BE}
\EF
The maximum length of the edge $\Delta l$ in 
these elements is 0.14. The condition that
the corresponding de Broglie wavelength $2 \pi/k$ is longer than 
the four times of
the interval of the boundary elements yields a rough estimate of
the validity condition of the calculation as $k\!<\!11$. On each
$\Gamma_J$, u and q $\equiv \del u/\del n$ are approximated by
low order polynomials. For simplicity, we use constant elements:
\begin{equation}
u(\x_J)=u^J=Const.\,,~~~~q(\x_J)=q^J=Const.\,,~~~ \mathrm{on}~~ \Gamma_J.\label{eq:qu}
\end{equation}  
Substituting Eq.(\ref{eq:qu}) into Eq.(\ref{eq:bem2d}), we obtain
\begin{eqnarray}
\sum_{J=1}^N H_{IJ} u^J &=& \sum_{J=1}^N G_{IJ} q^J,
\nonumber
\\
H_{IJ}&=&\left\{
\begin{array}{@{\,}ll}
\T{H}_{IJ} & \mbox{$I \neq J$}\\
\T{H}_{IJ}-\f{1}{2} & \mbox{$I=J$}, 
\nonumber
\end{array}
\right.
\end{eqnarray}
where 
\begin{equation}
\T{H}_{IJ}\equiv \int_{\Gamma_{J}}\f{\del G_k}{\del n}
(\x_I,\y_J) \,dS(\y_J),~~G_{IJ}\equiv  \int_{\Gamma_{J}}
G_k(\x_I,\y_J) \,dS(\y_J). \label{eq:bemIJ}
\end{equation}
The singular integration must be carried out
for I-I components as the fundamental solution diverges at
($\x_{I}=\y_{I}$). This is not an intractable problem. Several numerical 
techniques have already been proposed by some authors \cite{Tel,Hay}. 
We have applied Hayami's method 
to the  evaluation of  the singular integrals \cite{Hay}. Introducing
coordinates similar to spherical coordinates 
centered at $\x_I$, the singularity is 
canceled out by the Jacobian which makes the integral regular.    
\\
\indent
Let $g_i\,{(i\!=\!1,2,\ldots,8)}$ be the generators of the discrete 
group $\Gamma$ which identify a boundary face $F_i$ 
with another boundary face $g_i(F_i)$: 
\begin{equation}
g_i(\x_i)=\x_i,~~~~~~~~~~~ \x_i \in F_i.
\end{equation}
The boundary of the fundamental domain can be divided into two
regions $\del \Omega_A$ and $\del \Omega_B$ and each of them consists
of $N/2$ boundary elements,
\begin{equation}
 \del \Omega_A={\cup F_i},~~~~~~~~ \del \Omega_B=
{\cup g_i(F_i)},~~i=1,2, \ldots,8.
\end{equation}
The periodic
boundary conditions 
\begin{equation}
u(g_i(\x_i))=u(\x_i),~~~~q(g_i(\x_i))=-q(\x_i),~~~~~~i=1,2,\ldots ,8
\end{equation}   
reduce the number of the independent variables to N, \textit{i.e.}
for all $\x_B \in \del \Omega_B$, there exist $g_i \in \Gamma$ and
$\x_A \in \del \Omega_A$ such that 
\begin{equation}
u(\x_B)=u(g_i(\x_A))=u(\x_A),~~~~q(\x_B)=
-q(g_i(\x_A))=-q(\x_A).
\end{equation}  
Substituting the above relation into Eq.(\ref{eq:bemIJ}), we obtain
\begin{equation}
\left[
\begin{array}{@{\,}cc@{\,}}
H_{AA}+H_{AB} & -G_{AA}+G_{AB}\\ 
H_{BA}+H_{BB} & -G_{BA}+G_{BB } 
\end{array}
\right]
\left\{
\begin{array}{@{\,}c@{\,}}
u_A\\ 
q_A
\end{array}
\right\}
=0, \label{eq:uq}
\end{equation}
where $u_A=(u^1,u^2, \ldots u^{N/2})$ and $q_A=(q^1,q^2, \ldots
q^{N/2})$
and
matrices $H=\{H_{I J}\}$ and $G=\{G_{I J}\}$ are written as 
\begin{equation}
H=
\left[
\begin{array}{@{\,}cc@{\,}}
H_{AA}& H_{AB}\\ 
H_{BA}& H_{BB} 
\end{array}
\right],~~~
G=
\left[
\begin{array}{@{\,}cc@{\,}}
G_{AA}& G_{AB}\\ 
G_{BA}& G_{BB} 
\end{array}
\right].
\end{equation}
Eq. (\ref{eq:uq}) takes the form
\begin{equation}
[A(k)]\{x\}=0, \label{eq:Ax2}
\end{equation}
where N$\times$N-\,dimensional matrix $A$ is constructed from  $G$ and
$H$ and N-\,dimensional vector $x$ is constructed from $u_A$ and $q_A$.
For the presence of the non-trivial solution, the following relation 
must hold,
\begin{equation}
\mathrm{det}[A(k)]=0. \label{eq:detA}
\end{equation}
Thus the eigenvalues of the Laplace-Beltrami operator in a
CH space are obtained by
searching for $k$'s which satisfy Eq.(\ref{eq:detA}). 
In practice, Eq.(\ref{eq:detA}) cannot be exactly satisfied as the
test function which has a locally polynomial behavior is slightly
deviated from the exact eigenfunction. 
Instead, one must search for 
the local minima of \textrm{det}[A(k)]. This process needs long
computation time as $A(k)$ depends on k implicitly. Our numerical
result ($k\!<\!10$) is shown in table \ref{tab:eg}.
\begin{table}
\begin{center}
\begin{tabular}{cc}   \hline
\multicolumn{1}{c}{k} &
\multicolumn{1}{c}{$m_k$} 
\\ \hline
        5.41 & 1       \\ \hline
        5.79 & 1       \\ \hline   
        6.81 & 1    \\ \hline
        6.89 & 1    \\ \hline
        7.12 & 1    \\ \hline
        7.69 & 1    \\ \hline
        8.30 & 1      \\ \hline
        8.60 & 1     \\ \hline
        8.73 & 1    \\ \hline
        9.26 & 2     \\ \hline 
        9.76 & 1        \\ \hline
        9.91 & 1      \\ \hline
        9.99 & 1      \\ \hline
\end{tabular}\caption{Eigenvalue k and multiplicity $m_k$} 
\label{tab:eg}
\end{center} 
\end{table}
\\
\indent
The first ''excited state'' which corresponds to $k\!=\!k_1$ is 
important for the
understanding of CMB anisotropy. Our numerical result $k_1\!=\!5.41$
is consistent with the value $5.04$ obtained from Weyl's
asymptotic formula 
\begin{equation}
N[\nu]=\f{V\!o\,l ({\cal M})\nu^3}{6 \pi^2},~~\nu\equiv \sqrt{k^2-1},
~~~~\nu\!>\!>\!1,
\end{equation} 
assuming that no degeneracy occurs. 
One can interpret the first excited state as the mode that has 
the maximum de Broglie wavelength $2 \pi/k_1$. Because of the 
periodic boundary conditions, the de Broglie wavelength  
can be approximated by the 
''average diameter'' of the fundamental domain defined as a sum of
the inradius $r_-$ and the outradius $r_+$\footnote{The inradius 
$r_-$ is the radius of the largest
simply-connected sphere in the fundamental domain, and the outradius $r_+$
is the radius of the smallest sphere that can enclose the
fundamental domain. $r_-\!=\!0.535$, $r_+\!=\!0.7485$ for
the Thurston manifold.},
which yields $k_1=4.9$ just 
$10\!\%$ less than the numerical value. From these estimates,
supercurvature modes in small CH spaces ($Vol({\cal M})\sim 1$) 
are unlikely to be observed.
\\
\indent
To compute the value of eigenfunctions inside the fundamental domain,
one needs to solve Eq.(\ref{eq:Ax2}). The singular decomposition
method is the most suitable numerical method for solving any linear
equation with a singular matrix $A$, which can be 
decomposed as 
\begin{equation}
A=U^{\dagger}D V,
\end{equation}
where $U$ and $V$ are unitary matrices and D is a diagonal matrix. If 
$D_{ii}$ in $D$ is almost zero then the complex conjugate of the i-th
row in V is an approximated solution of Eq.(\ref{eq:Ax2}).    
The number of the ''almost zero'' diagonal elements in D is equal to 
the multiplicity number.  
\BF
\IG[width=15cm]{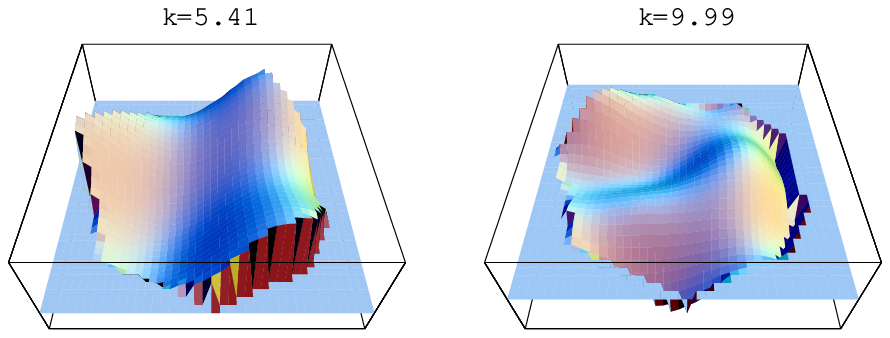}
\caption{Eigenfunctions $u_\nu$ on a slice $x_3=0.0087$}
\label{fig:EF}
\EF
Substituting the values of the eigenfunctions and their normal 
derivatives on the
boundary into Eq.(\ref{eq:inter}), the values of the 
eigenfunctions inside the fundamental domain can be computed. 
Eigenfunctions $k\!=\!5.41$ and $k\!=\!9.99$
in Poincar$\acute{\textrm{e}}$
coordinates plotted as ($x_1,x_2,h$), where $h\!=\!u_\nu
(x_1,x_2,0.0087)$ 
are shown in 
figure \ref{fig:EF}. The eigenfuctions we computed are all
real-valued. Note that the
non-degenerated eigenfunctions must be real-valued.  
\\
\indent
The numerical accuracy of the obtained eigenvalues is roughly estimated as
follows. First, let us write the obtained numerical solution in terms
of the exact solution as $k=k_0+\delta k$ 
and $u_k(\x)=u_{k_0}(\x)+\delta u_k(\x)$,
where $k_0$ and $u_{k_0}(\x)$ are the exact eigenvalue and
eigenfunction, respectively. The singular decomposition method
enables us to find the best approximated solution which satisfies
\BE
[A]\{x\}=\epsilon, ~~~|\epsilon | <<1, 
\label{eq:epsilon}
\EE
where $\epsilon$ is a N-dimensional vector and $|~ |$ denotes
the Euclidean norm. It is expected that 
the better approximation gives the smaller $|\epsilon |$.
Then Eq. (\ref{eq:epsilon}) can be written as, 
\BE
\int_{\Omega} G_{k_0+\delta k}(\x,\y_J) (\Delta+(k_0+\delta k)^2)
(u_{k_0}(\x)+\delta u_k(\x))\,\sqrt{g}\, dV_\x=\epsilon(\y_J).
\label{eq:eps}
\EE
Ignoring the terms in second order, Eq.(\ref{eq:eps}) is reduced to 
\BE
\int_{\Omega} G_{k}(\x,\y_J)((\Delta+k_0^2)\delta u_k(\x)
+2 k \delta k u_{k}(\x))\,\sqrt{g}\, dV_\x=\epsilon(\y_J).
\EE
Since it is not unlikely that $(\Delta+k_0^2)\delta u_k(\x)$
is anticorrelated to $2 k \delta k u_{k}(\x)$, 
we obtain the following relation by averaging over $\y_J$,
\BE
2 k |\delta k| \Biggl <\biggl|\int_{\Omega} G_{k}(\x,\y_J)u_{k}(\x)\,
\sqrt{g}\, dV_\x \biggr|\Biggr> \sim\,\,<|\epsilon|>,
\EE
where $< >$ denotes the averaging over $\y_J$  
Thus one can estimate the expected deviation of 
the calculated eigenvalue $|\delta k|$ from $u_k(\x)$ and $\epsilon(\y_J)$.  
We numerically find that $|\delta k|=0.005$ for $k=5.41$ 
and $|\delta k|=0.01$ for
$k=9.91$. The other deviation values lie in between $0.005$ and
$0.01$. 
\\
\indent
By computing the second derivatives, one
can also estimate the accuracy of the computed eigenfunctions. 
The accuracy parameter $err$ is defined as
\BE
err(k,\x)\equiv (\Delta + k^2) u_k(\x),
\EE
\BF
\IG[width=15cm]{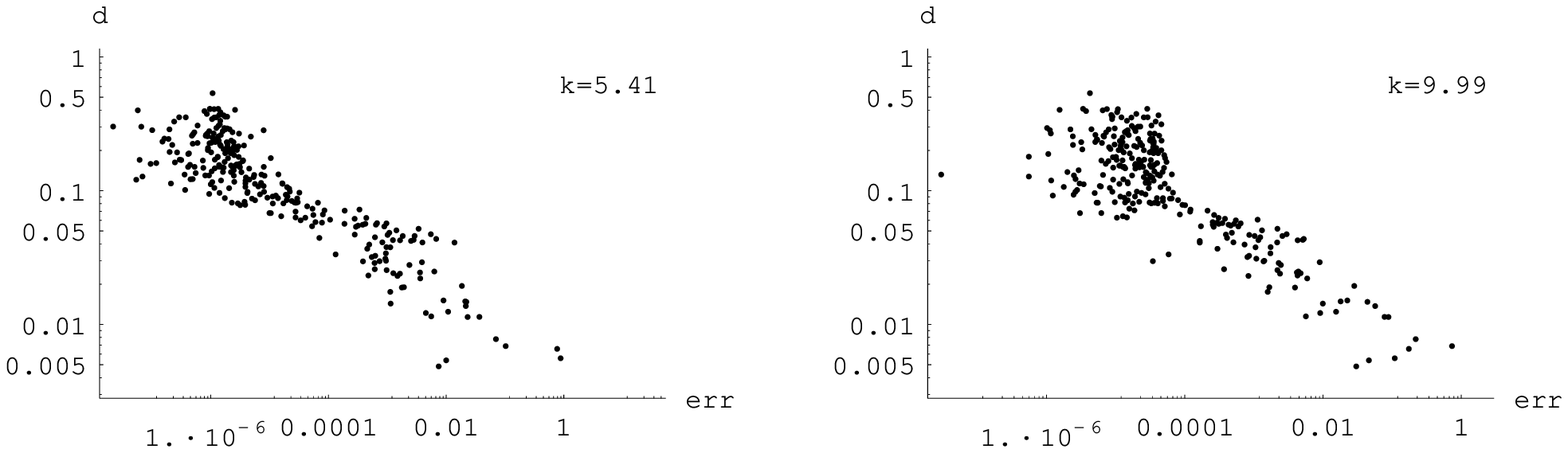}
\caption{Error ($err$) versus hyperbolic distance to the boundary for 291
points inside the fundaental domain. d denotes
the hyperbolic distance from the evaluation point $\x$ to the 
nearest point on the
boundary. }
\label{fig:err}
\EF
where $u_k(\x)$ is normalized (${\cal O}(u_k(\x))\sim 1$). We see from figure 
\ref{fig:err} that the accuracy becomes worse
as the evaluation point approaches the boundary. However, for points with 
hyperbolic distance $d>0.1$ between the evaluating point and the
nearest boundary, the errors are very small 
indeed: $err\lesssim
10^{-4\sim -5}$. This result is considered to be 
natural because the characteristic 
scale $L$ of the boundary elements is $\sim 0.07$ for our 1168 elements. 
If $d<L$, the integrands in Eq.(\ref{eq:inter}) become appreciable on
the neighborhood of the nearest boundary
point because the free Green's
function approximately diverges on the point. 
In this case, the effect of the deviation from the exact 
eigenfunction is significant. If $d>>L$, the integrand on 
all the boundary points contributes almost equally to the 
integration so that the local deviations are cancelled out.  
\\
\indent
As we shall see in the next section, expansion coefficients
are calculated using the values of eigenfunctions on a sphere. 
Since the number of evaluating points which
are very close to the boundary is negligible on the sphere, 
expansion coefficients can be computed with
relatively high accuracy.
\pagebreak
\section{Statistical properties of eigenmodes} 
\indent

Properties of eigenmodes of the Laplace-Beltrami operator
are determined by the Helmholtz 
equation. Therefore, at first glance it does not seem to make a sense 
to study the statistical properties of the eigenmodes. 
\\
\indent
However, if one recognizes the Laplace-Beltrami operator in
a CH space as the Hamiltonian in a quantum mechanical system, 
each eigenmode can be interpreted as a wavefunction in an eigenstate. 
Since the corresponding classical
system is known to be a typical chaotic system (K-system), it is
natural to assume that the imprints of classical chaos is hidden in
the corresponding quantum system.
\\
\indent
Recent studies have demonstrated that some of the statistical 
properties of energy spectrum are in accordance with the universal 
prediction of random-matrix theory(RMT) \cite{Meh,Boh}. 
RMT also predicts that the 
squared expansion coefficients $|a_i|^2$ of an eigenstate with respect to 
a generic basis are Gaussian distributed \cite{Aur1,Haake,Bro}. 
In the limit $N\rightarrow \infty$, $x=|a_i|^2$ obeys the statistics 
given for three universality classes of the orthogonal (GOE,~$\mu=1$), unitary
(GUE,~$\mu=2$) and symplectic (GSE,~$\mu=4$) ensembles, each
distribution function $P$ is given by
\begin{equation}
P_{\mu}(x)= \Biggl(\f{\mu}{2}\Biggr) ^{\mu/2}~
\Gamma(\mu/2)~x^{\mu/2-1}~e^{-\mu x /2}.
\label{eq:dis}
\end{equation}
In our case, as the time-reversal symmetry of the Hamiltonian implies,
one expects that $|a_i|^2$ obeys the GOE prediction.
In order to apply the GOE prediction to the statistical properties of 
eigenstates on CH spaces, one needs to find a set of orthonormal
basis but no closed analytic expression is known for any CH spaces.
To avoid the problem, Aurich and Steiner noticed  that the  
wavefunctions on the hyperbolic octagons can be continued onto the 
universal covering space $\M H^2$, 
and eigenstates can be expanded in terms of circular-waves
\cite{Aur1}. 
They numerically found that the squared expansion coefficients 
obeys the GOE prediction in highly excited states of
a hyperbolic asymmetrical octagon model.  
\\
\indent
We extend their method to three-dimensional CH models where we consider only
low-lying modes. First, we normalize the obtained 14 eigenfunctions 
on the Thurston manifold. The eigenfunctions are naturally continued 
onto the whole 
unit Poincar$\acute{\textrm{e}}$ ball by the periodic boundary condition. 
As a ''generic basis'', we consider a set of orthonormal
eigenfunctions $Q_{\nu lm}$ (T-complete functions) on the unit 
pseudosphere which is isometric to the Poincar$\acute{\textrm{e}}$ ball, 
\begin{eqnarray}
Q_{\nu lm} &\equiv& X_{\nu l}(\chi) Y_{lm}(\theta,\phi),
\nonumber
\\
\nonumber
\\
X_{\nu l}
&\equiv&(-1)^{l+1} \sqrt{\f{2}{\pi}} 
\Biggr(\prod_{n=0}^l(n^2+\nu^2)\Biggl)^{-1/2} \sinh^l \chi 
\f{d^{l+1}(\cos \nu \chi)}{d(\cosh \chi)^{l+1}},  
\nonumber
\\
\nonumber
\\
&=&\f{\Gamma  (l+1+\nu i)}{\Gamma (\nu i)} \sqrt{\f{1}{\sinh \chi}}
P^{-l-1/2}_{\nu i-1/2}(\cosh
\chi),~~~~\nu^2=k^2-1,
\end{eqnarray}
where $P$, $Y_{lm}$ and  $\Gamma$ denote the associated Legendre 
function, the spherical harmonics and gamma function, respectively.
 $P$ can be written in 
terms of the hypergeometric function $_2{\cal F}_1$ \cite{Harrison},
\begin{equation}
P^{-l-1/2}_{\nu i-1/2}(\cosh\chi)=\f{\bigl(\coth \f{1}{2} \chi \bigr)
^{-l-1/2}}{\Gamma\bigl(\f{3}{2}+l\bigr)}\,\,{}_2{\cal F}_1
\Biggl(\f{1}{2}-\nu i,
\f{1}{2}+\nu i;\f{3}{2}+l;-\sinh^2\f{1}{2} \chi \Biggr).
\end{equation}
Eigenfunctions $u_\nu$ can be expanded in terms of $Q_{\nu l m}$'s as
\begin{equation}
u_\nu=\sum_{l m} \xi_{\nu l m}\,X_{\nu l}(\chi) Y_{l m}(\theta,\phi).
\label{eq:ux}
\end{equation}
Note that each $u_\nu$ has no components with 
$\nu^{'} \neq \nu $ because $Q_{\nu l m}$'s are 
complete and linearly independent. 
\\
\indent
At first glance the computation of $\xi_{\nu l m}$ in 
Eq. (\ref{eq:ux}) seems cumbersome as the domain of the integration 
extends over the whole pseudosphere. 
Fortunately, one can obtain $\xi_{\nu l m}$ by evaluating  
two-dimensional integrals.  $\xi_{\nu l m}$ can be written as
\begin{equation}
\xi_{\nu l m} X_{\nu l}(\chi_o)=\int u_\nu(\chi_o,\theta,\phi)\, 
Y^*_{l m}(\theta,\phi) d \Omega, \label{eq:Intxi}
\end{equation}   
which is satisfied for the arbitrary value of $\chi_o$. 
In practice the numerical
instability occurs in the region where the
absolute value of $X_{\nu l}$ is too small. In our computation, 
the values of $\chi_o$ are chosen as shown in table \ref{tab:PACHI}.
\begin{table}
\begin{center}
\begin{tabular}{|c|c|c|c|}   \hline
\multicolumn{1}{|c|}{ } &
\multicolumn{1}{|c|}{0 $\leq$\,l\,$<$\,8\,} &
\multicolumn{1}{|c|}{8 $\leq$ l $\leq$ 13} &
\multicolumn{1}{|c|}{13 $<$ l $\leq$ 18}  \\ \hline
    k$<$8   &   0.53 & 1.3 & 1.6        \\
    k$>$8   &   0.53 & 1.1 & 1.3         \\ \hline   
\end{tabular}\caption{An example of choice of $\chi_o$
for which the absolute value of 
$X_{\nu l}(\chi_o)$ is not too small.} 
\label{tab:PACHI}
\end{center} 
\end{table}
Thus $\xi_{\nu
l m}$ can be computed if one obtains the values of eigenfunctions
on the sphere 
\begin{equation}
x_1=\tanh\f{\chi_o}{2} \sin \theta \cos \phi,~x_2= 
\tanh\f{\chi_o}{2} \sin \theta \sin \phi,~x_3= \tanh\f{\chi_o}{2}
 \cos \theta 
\end{equation}
with radius $\chi_o$.  
\\
\indent
In order to compute the values of eigenfunctions on the sphere
with radius longer than the inradius $r_-\!=\!0.535$,
the points outside of the fundamental domain must be pulled back to the
inside, since Eq.(\ref{eq:inter}) is valid only if $\y$ is a set of
coordinates of an internal point. 
\\
\indent
The plots of eigenfunctions 
on a sphere $\chi_o\!=\!1.6$ are shown in figure \ref{fig:EF541} and  
figure \ref{fig:EF999}. Apparent structure of the eigenfunctions on the sphere 
seems complicated. However, some regular patterns are hidden in the
structure due to the periodic boundary conditions. Actually, there
are pairs of highly
correlated points on the sphere, since
any partial surface $S_{i1}$ of the sphere that is enclosed by copies
of the boundary of the fundamental domain pulled back to 
inside the fundamental
domain by the corresponding element of the discrete isometry group 
intersects another partial surface $S_{i2}$ that is pulled
back to inside the fundamental domain. 
To evaluate the correlation pattern, let us 
estimate how often a sphere with radius
$\chi=\chi_o$ intersects the copies of the fundamental domain.      
The approximate number $n_1$ of the copies of the 
fundamental domain inside
the sphere with radius (in proper length) $\chi_o$ is given by
\begin{equation}
n_1=\f{\pi (\sinh(2 \chi_o)-2 \chi_o)}{V\!o\,l(Q_2)}.
\end{equation}   
From this formula, in the case of the Thurston manifold, $n_1\!\sim\!29$ if
$\chi_o\!=\!1.6$. Because the sphere intersects the fundamental domain 
at random, the copies of the fundamental domain on the sphere
stick out their half portions on average. Therefore, the approximate 
number $n_2$ of the copies that intersect the sphere is given by
\begin{equation}
n_2=\f{\pi (\sinh(2(\chi_o+r_{ave}))-\sinh(2(\chi_o-r_{ave}))-
4 r_{ave})}
{V\!o\,l(Q_2)},
\end{equation}
where $r_{ave}=(r_++r_-)/2$.
This estimate gives  $n_2 \!\sim\! 120$ if 
$\chi_o\!=\!1.6$.  Approximating each eigenmode by de Broglie
waves, we obtain the corresponding fluctuation scale $\delta \!A$ in 
steradian on the sphere,
\begin{eqnarray}
\delta A
&=&\f{4 \pi}{n_2}\, \f{4 \pi^2}{k^2}
\nonumber
\\
&=&\f{16 \pi^2~V\!o\,l(Q_2)}  {k^2 (\sinh(2(\chi_o+r_{ave}))-\sinh(2(\chi_o-r_{ave}))-
4 r_{ave})}. \label{eq:FS} 
\end{eqnarray}   
Thus correlation patterns are observed in pairs of patches with 
typical size $\delta \! A$.   
When $\chi_o\!=\!1.6$, angular fluctuation scales are given
as $\delta l\!\sim\!\sqrt{\delta A}=\!21^o,~12^o$, for 
$k=5.41,\,9.99$, respectively.  
\BF
\IG[width=15cm,clip]{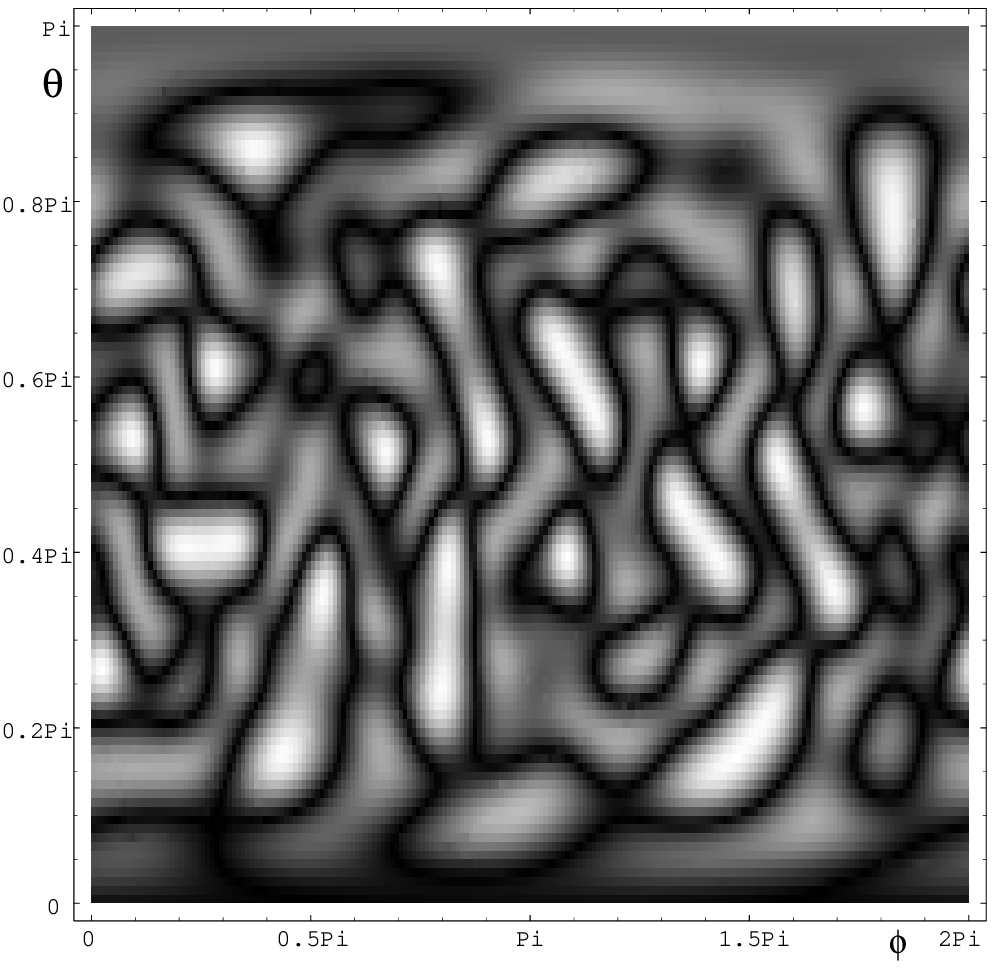}
\caption{Absolute values of eigenfunction $u(k=5.41)$ on 
sphere $\chi_o=1.6$ }
\label{fig:EF541}
\EF
\BF
\IG[width=15cm,clip]{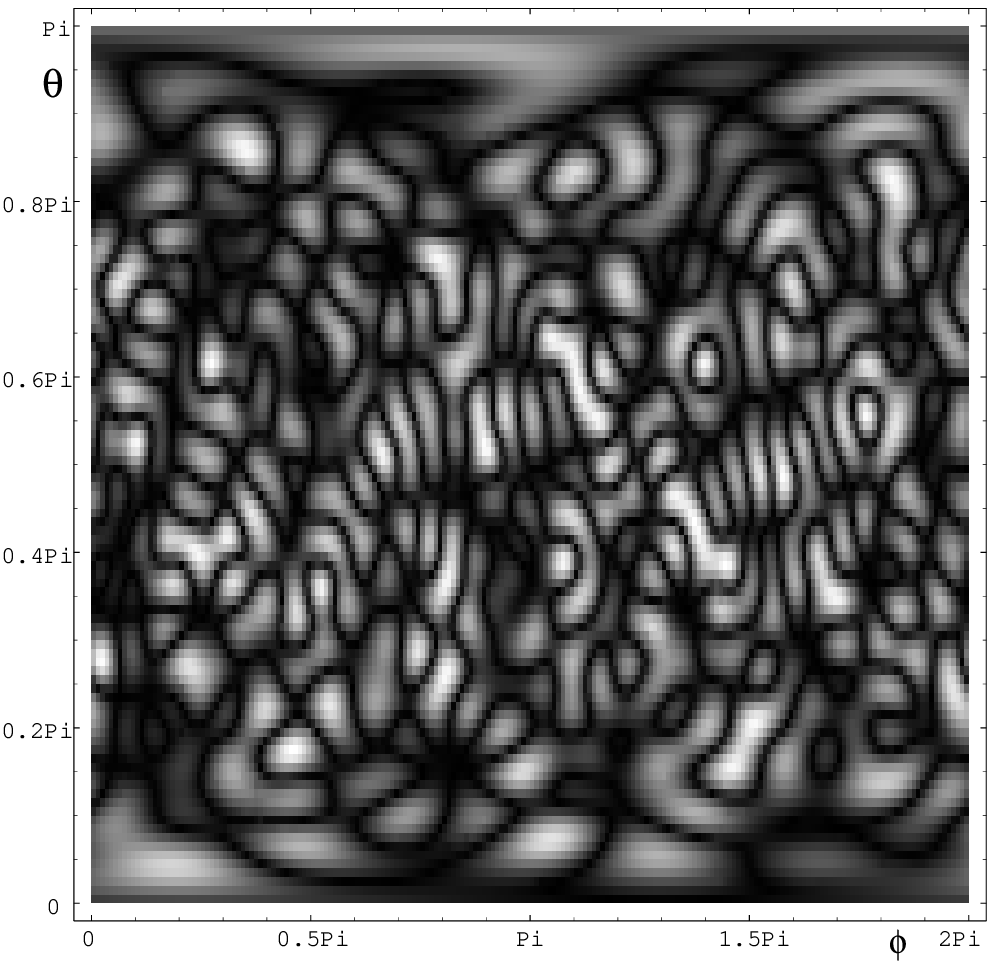}
\caption{Absolute values of eigenfunction $u(k=9.99)$ on 
sphere $\chi_o=1.6$}
\label{fig:EF999}
\EF
\\
\indent
Next, we extract a set of independent variables from $\xi_{\nu l m}$ 's.  
In general, any $Q_{\nu l m}$ is related to $Q_{\nu l -m}$  as
\begin{equation}
Q_{\nu \,l -m}=(-1)^m Q^*_{\nu l m} F(\nu, l),
\label{eq:Q}
\end{equation}  
where
\begin{equation}
F(\nu, l)=\f{\Gamma(l+\nu i+1)}{\Gamma(\nu i)}
\f{\Gamma(-\nu i)}{\Gamma(l-\nu i+1)}.
\end{equation}
If $u_\nu$ is real, from Eq.(\ref{eq:Q}),
\begin{eqnarray}
\nonumber
u_\nu
&=&\sum_{l\, m} \xi_{\nu l m}\,Q_{\nu l m}
\\
\nonumber
&=&\sum_{l\, m} \xi^*_{\nu l m}\,Q^*_{\nu l m}
\\
&=&\sum_{l\, m} (-1)^{-m}\, \xi^*_{\nu\, l -m}\,Q_{\nu l m}F^{-1}(\nu,l),
\label{eq:exp}
\end{eqnarray}
therefore,
\begin{equation}
\xi_{\nu l m}=(-1)^m\, \xi^*_{\nu l -m}\,F^{-1}(\nu,l).
\label{eq:xi}
\end{equation}
Thus $\xi_{\nu l -m}$ can be written in terms of $\xi_{\nu l m}$.
To extract a set of independent variables from $\xi_{\nu l m}$'s,
we rewrite Eq.(\ref{eq:exp}) as follows
\begin{eqnarray}
\nonumber
u_\nu
&=&\sum_{l,\,  m\leq 0} \xi_{\nu l m}\,Q_{\nu l m}+
\sum_{l,\, m > 0} \xi^*_{\nu l m}\,Q^*_{\nu l m}
\\
\nonumber
&=&-\textrm{Im}(\xi_{\nu 0 0})\, \textrm{Im}(Q_{\nu 0 0})+
\sum_{l>0} \sqrt{c_{\nu l}} \textrm{Re}(\xi_{\nu l 0})R_{\nu l 0}
\\
&+&
\sum_{l>0\,, m>0} \sqrt{2}\Biggl( \textrm{Re}(\xi_{\nu l m})
\sqrt{2} \textrm{Re}(Q_{\nu l m})-
\textrm{Im}(\xi_{\nu l m})
\sqrt{2}\, \textrm{Im}(Q_{\nu l m}) \Biggr),
\end{eqnarray}
where
\begin{equation}
R_{\nu l 0}=(c_{\nu l})^{-1/2}\Biggl(\textrm{Re}(Q_{\nu l 0})+
\f{1-F(\nu,l)}{1+F(\nu,l)} 
\textrm{Im}(Q_{\nu l 0})i \Biggr ),
\end{equation}
and
\begin{eqnarray}
\nonumber
c_{\nu l}&=&1- \f{(1-F(\nu,l))^2}{2 F(\nu,l)}-
\Biggl\{ \biggl(\f{1-F(\nu,l)}{1+F(\nu,l)}\biggr)^2+1 \Biggr\}
\\
\nonumber
&\times&
\textrm{Im} \Biggl( \f{\Gamma(l+\nu i+1)}{\Gamma(\nu i)}\Biggr)^2 \Biggl |
\f{\Gamma(\nu i)}{\Gamma(l+\nu i+1)} \Bigg|^2
\\
&=&\f{2}{1+\textrm{Re}(F(\nu,l))}.
\end{eqnarray}
Thus the real eigenfuctions can be written in terms of real
independent coefficients $a_{\nu l m}$ and real-valued
$R_{\nu l m}$,
\begin{equation}
u_\nu=\sum_{l,\, m} a_{\nu l m}\,R_{\nu l m},
\end{equation}
where
\begin{eqnarray}
\nonumber
a_{\nu 0  0}\!\!&=&\!\!-\textrm{Im}(\xi_{\nu 0  0}),~~~~
a_{\nu l  0}=\sqrt{c_{\nu l}}\textrm{Re}(\xi_{\nu l 0}),~~~~l>0,
\\
a_{\nu l  m}\!\!&=&\!\!\sqrt{2}\textrm{Re}(\xi_{\nu l m}),~~m>0,
~~~~a_{\nu l  m}=-\sqrt{2}\textrm{Im}(\xi_{\nu l -m}),~~m<0,
\end{eqnarray}
and
\begin{eqnarray}
\nonumber
R_{\nu 0  0}\!\!&=&\!\!\textrm{Im}(Q_{\nu 0  0}),
\\
R_{\nu l  m}\!\!&=&\!\!\sqrt{2}\textrm{Re}(Q_{\nu l m}),~~m>0,~~~
R_{\nu l  m}=\sqrt{2}\textrm{Im}(Q_{\nu l -m}),~~m<0.
\end{eqnarray}
\\
\indent
Now we turn to the statistical properties of the coefficients $a_{\nu
l m}$.  As in \cite{Aur1}, we consider the 
cumulative distribution of following quantities, 
\begin{equation}
\f{|a_{\nu l m}-\bar{a}_{\nu}|^2}{\sigma_{\nu}^2}
\end{equation}
where $\bar{a}_{\nu}$ is the mean of $a_{\nu
l m}$'s and $\sigma_{\nu}^2$ is the variance. The cumulative
distribution  is compared  
to the cumulative RMT distribution functions which are directly derived 
from Eq.(\ref{eq:dis}),
\begin{eqnarray}
\nonumber
I_\mu(x)&=&\int_0^x d x\acute{ }\, P_{\mu}(x\acute{ }\,)
\\
&=&\f{\gamma (\mu/2,\mu x/2 )}{\Gamma(\mu/2)},
\end{eqnarray} 
where $\gamma(x,y)$ is the incomplete gamma function. To test the
goodness of fit between the computed cumulative distribution function
 and that predicted by RMT, we use Kolmogorov-Smirnov statistic $D_N$ which
is the least upper bound of all pointwise differences $|I_N(x)-I(x)|$
 \cite{Hog},
\begin{equation}
D_N\equiv \sup_{x} |I_N(x)-I(x)|,
\end{equation}
where $I_N(x)$ is the empirical cumulative distribution function
defined by
\begin{eqnarray}
 I_N(x)&=&\left\{ \begin{array}{@{\,}ll}
0, & x<y_1,
\\
j/N,~~& y_j \leq x < y_{j+1},~~~~j=1,2,\ldots,N\!-\!1,
\\
1, & y_j \leq x, 
\end{array}
\right. 
\end{eqnarray}
where $y_1<y_2< \ldots <y_N$ are the computed values of a random
sample which consists of $N$ elements. 
If $I_N(x)$ is ''close'' to $I(x)$, the observed $D_N$ must be so
small that it falls within the range of possible fluctuations 
which depend on the size of the random sample.
For the random variable $D_N$ for any $z>0$, it can be shown that 
the probability of $D_N<d$ is given by \cite{Bir}
\begin{equation}
\lim_{N \rightarrow \infty} ~P(D_N<d=z N^{-1/2})=L(z),\label{eq:P} 
\end{equation}  
where
\begin{equation}
L(z)=1-2 \sum_{j=1}^{\infty} (-1)^{j-1} e^{-2j^2 z^2}.
\end{equation}
From observed maximum difference $D_N=d$, we obtain 
the significant level $\alpha_D\!=\!1-P$ which is equal to the probability 
of $D_N>d$. If $\alpha_D$ is found to be large enough, 
the hypothesis $I_N(x)=I(x)$ is verified. The 
computed cumulative distributions of $|a_{\nu l m}|^2$ and the
GOE($\mu=1$) prediction $I_1(x)$ for four examples are plotted 
in figure \ref{fig:KS},  
and the maximum difference $d$ and the significant levels $\alpha_D$ for
$0 \leq l \leq 10$ and  $0 \leq l \leq 18$ are
shown in table \ref{tab:KS}. Note that the last digit in $\alpha_N$
is not guaranteed, since Eq.(\ref{eq:P}) is an asymptotic formula. 
\\
\indent
We see from figure \ref{fig:KS} and 
table \ref{tab:KS} that the agreement with GOE prediction is
remarkably good. The Gaussian behavior for highly excited states 
is naturally expected as the semiclassical wavefunctions must reflect 
the chaotic nature in the corresponding classical systems. However, 
the Gaussian behavior for low-lying modes is not apparent at all. 
It is possible that the
non-gaussianity behavior is rather prominent as these modes have 
fluctuations on large scale and that reflects the   
pure quantum mechanical 
behavior\footnote{The typical scale of angular fluctuation of $l$ mode 
is approximately $\pi/(2l)$ and the typical scale of radial fluctuation  of
$k$ mode is approximately given by
$2 \pi/k$.}. Nevertheless, our numerical results serve to strengthen 
the hypothesis that the expansion coefficients behave as Gaussian 
pseudo-random numbers even for low-lying modes.   
\BF
\IG[width=15cm,clip]{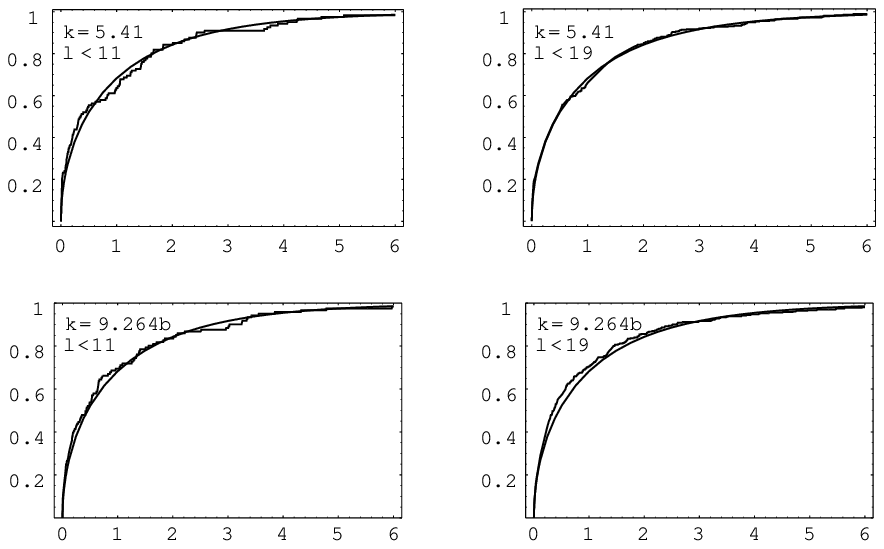}
\caption{The cumulative distribution of 
$|a_{\nu l m}-\bar{a}_{\nu l m}|^2/\sigma^2$ are plotted with the
theoretical curve for GOE prediction (full curve). The statistics are
based on 121 expansion coefficients  for $l<11$ and 361
coefficients for $l<19$.}
\label{fig:KS}
\EF
\begin{table}
\begin{center}
\begin{tabular}{c|cc|cc}   \hline
\multicolumn{1}{c}{}&
\multicolumn{2}{|c|}{$0 \leq l \leq\!10$}&
\multicolumn{2}{c}{$0 \leq l \leq\!18$}\\ \hline
\multicolumn{1}{c|}{k} &
\multicolumn{1}{c}{$D_N\times 10^{2}$} &
\multicolumn{1}{c|}{$\alpha_D$} &
\multicolumn{1}{c}{$D_N \times10^{2}$} &
\multicolumn{1}{c}{$\alpha_D$}  \\ \hline
        5.41  & 9.42  & 23.4\% & 4.29 & 51.8\% \\      
        5.79  & 3.88  & 99.3\% & 4.29 & 52.1\% \\         
        6.81  & 5.95  & 78.6\% & 3.47 & 77.8\% \\   
        6.89  & 9.34  & 24.1\% & 2.36 & 98.8\% \\  
        7.12  & 7.12  & 57.2\% & 2.59 & 96.8\% \\ 
        7.69  & 10.23~\, & 15.9\% & 4.84 & 36.7\% \\ 
        8.30  & 6.38  & 70.8\% & 3.88 & 65.0\% \\
        8.60  & 5.63  & 83.8\% & 2.09 & 99.7\% \\ 
        8.73  & 9.46  & 22.9\% & 2.78 & 94.3\% \\
        9.26a & 7.21  & 55.6\% & 3.46 & 78.1\% \\  
        9.26b & 5.99  & 77.9\% & 6.11 & 13.5\% \\ 
        9.76  & 7.41  & 52.0\% & 4.57 & 43.8\% \\ 
        9.91  & 8.90  & 29.3\% & 3.23 & 84.7\% \\ 
        9.99  & 4.27  & 98.0\% & 3.72 & 70.0\% \\ \hline 
\multicolumn{1}{c|}{ave.} &
\multicolumn{1}{c}{7.23} &
\multicolumn{1}{c|}{56.3\%} &
\multicolumn{1}{c}{3.69} &
\multicolumn{1}{c}{68.8\%} \\ \hline
\end{tabular}\caption{The Kolmogolov-Smirnov statistics $D_N$ and the
significance levels $\alpha_D$ for the test of the hypothesis
 $I_N(x) \neq I(x)$ and their averages. $N=121$ for 
$0 \leq l \leq\!10$ and $N=361$ for
$0 \leq l \leq\!18$. The mode $k\!=\!9.26$ is degenerated into two modes,
which (after orthogonalization)are denoted by $k\!=\!9.26a$  and 
$k\!=\!9.26b$. } 
\label{tab:KS}
\end{center} 
\end{table}
\pagebreak
\\
\indent
Next we examine the randomness of $a_{\nu l m}$'s. Because $a_{\nu l
m}$'s are actually determined by the Helmholtz equation, it is appropriate
to describe $a_{\nu lm}$'s as pseudo-random numbers. We apply the 
run test for testing randomness (see \cite{Hog}). 
\\
\indent
Suppose that we have $n$ observations
of the random variable $X$ and m observations of the random variable
$Y$. The combination of those variables into $m\!+\!n$ observations
placed in ascending order of magnitude yields
\begin{center}
\large\textit{
\U{xxx} \U{yy} \U{xx} \U{yyy} \U{x} \U{y} \U{xx} \U{yy}},
\end{center}
where $x$ denotes an observation of $X$ and $y$ denotes an observation 
of $Y$. Each underlined group which consists of successive 
values of $X$ or $Y$ is called \textit{run}.  The statistics of 
number of runs are used for testing whether $X$ and $Y$ have 
the same the distribution function. Regardless of the type of the distribution
function, the run number $r$ is known to behave as Gaussian random
numbers in the limit $m,\,n \rightarrow \infty$.
\\
\indent
The run test is also used as a test
for randomness. Let $a_1,a_2,\ldots, a_N$ be the observed values of a
random variable $A$. For simplicity, assume that $N$ is even. The
median divides the observed values into a lower and an upper half.
It is represented as $L$ if it falls below the median, and it is 
represented as $U$ if it falls above the median.  For instance, a
sequence
\begin{center}
\textit{
\U{UUU} \U{L} \U{UU} \U{LL} \U{U} \U{U} \U{L} \U{UU}},
\end{center}
has 8 numbers of runs ($r\!=\!8$). The critical region for testing 
the hypothesis of randomness is of the form $r<c_1$ or $r>c_2$ where 
$c_1$ and $c_2$ is
readily given by the Gaussian distribution function. The significant
level $\alpha_r$ is the probability of $r>c_1$ or $r>c_2$. As the
Kolmogolov-Smirnov test, $\alpha_r$ is given by the observed $r$.
\\
\indent
The run numbers $r$ and the significant levels $\alpha_r$ are shown
in table \ref{tab:R}. High significant levels are again observed except
for the one at $k\!=\!8.73$ for $0 \leq l \leq\!18$. As the
corresponding $r$ is larger than the averaged value, this may be due
to the cyclic effect. On the whole, it is 
concluded that $a_{\nu l m}$ 's behave as if they are random variables. 
\\
\indent
   
\begin{table}
\begin{center}
\begin{tabular}{c|cc|cc}   \hline
\multicolumn{1}{c}{}&
\multicolumn{2}{|c|}{$0 \leq l \leq\!10$}&
\multicolumn{2}{c}{$0 \leq l \leq\!18$}\\ \hline
\multicolumn{1}{c|}{k} &
\multicolumn{1}{c}{$r$} &
\multicolumn{1}{c|}{$\alpha_r$} &
\multicolumn{1}{c}{$r$} &
\multicolumn{1}{c}{$\alpha_r$}  \\ \hline
        5.41  & 62  & 85.5\% & 185 & 67.3\% \\      
        5.79  & 58  & 46.5\% & 174 & 39.9\% \\         
        6.81  & 69  & 14.4\% & 196 & 11.4\% \\   
        6.89  & 60  & 71.5\% & 168 & 14.0\% \\  
        7.12  & 69  & 14.4\% & 184 & 75.2\% \\ 
        7.69  & 57  & 36.1\% & 191 & 29.2\% \\ 
        8.30  & 63  & 71.5\% & 177 & 59.8\% \\
        8.60  & 59  & 58.3\% & 184 & 75.2\% \\ 
        8.73  & 70  & 10.0\% & 201 & 3.5\% \\
        9.26a & 56  & 27.3\% & 177 & 59.8\% \\  
        9.26b & 55  & 20.1\% & 182 & 91.6\% \\ 
        9.76  & 59  & 58.3\% & 182 & 91.6\% \\ 
        9.91  & 70  & 10.0\% & 196 & 11.4\% \\ 
        9.99  & 58  & 46.5\% & 179 & 75.2\% \\ \hline 
\multicolumn{1}{c|}{ave.} &
\multicolumn{1}{c}{61.8} &
\multicolumn{1}{c|}{40.7\%} &
\multicolumn{1}{c}{184} &
\multicolumn{1}{c}{50.4\%} \\ \hline
\end{tabular}\caption{The run numbers $r$ and the
significance levels $\alpha_r$ for the test of the hypothesis
that $a_{\nu l m}$ 's are random variables. $N=121$ for 
$0 \leq l \leq\!10$ and $N=361$ for
$0 \leq l \leq\!18$. The mode $k\!=\!9.26$ is degenerated into two modes,
which (after orthogonalization)are denoted by $k\!=\!9.26a$  and 
$k\!=\!9.26b$. } 
\label{tab:R}
\end{center} 
\end{table}
\section{Summary}

\indent
\\
\indent
In this paper, we have demonstrated that the DBEM is eminently suitable for 
computing eigenmodes on CH spaces and we obtain some low-lying eigenmodes on 
a CH space called Thurston manifold which is the second smallest 
in the known CH manifolds and we have studied the statistical properties of 
these eigenmodes.
\\
\indent
The low-lying eigenmodes are expanded in terms of eigenmodes on the
pseudosphere, and we find that the expansion coefficients behave as 
if they are Gaussian random numbers. Why are they so random even for 
low-lying modes? It should be pointed out that the 
randomness of the expansion coefficients for low-lying eigenmodes is
not  the property of the eigenmodes themselves but rather the property of the
images of eigenmodes on the whole universal covering space, since the
fluctuation scales for low-lying eigenmodes are comparable to the
the size of the fundamental domain. We conjecture that the origin of
the random behavior
of eigenmodes comes from the almost randomly 
distributed images of a set of points in the universal covering
space.  
\\
\indent
Computation of eigenmodes is essential in simulating the CMB in
CH cosmological models. As the DBEM needs only a set of face-to-face
identification maps and the discretization of the corresponding
fundamental domain, it can be applied to other CH spaces 
straightforwardly. However, the computation of the modes with small
fluctuation scale $k\!>\!>\!1$ is still a difficult task as the 
number of modes increases as $N\propto k^3$. 
\\
\indent
Nevertheless, the contribution of the modes
with small fluctuation to the temperature correlation of 
CMB can be estimated by assuming that
the expansion coefficients for excited states ($k\!>\!>1$) 
also behave as Gaussian
pseudo-random numbers as well as that for low-lying modes. 
The assumption is numerically 
confirmed in a two-dimensional CH model \cite{Aur1}.  
\\
\indent
If the observed Gaussian
pseudo-randomness is found to be the universal behavior in CH spaces
for low-lying modes as well as excited modes, 
the origin of the gaussianity of the CMB fluctuations can be partially 
explained. This is because the amplitude of the CMB fluctuation is written in
terms of:\\
1. expansion coefficients of the initial fluctuation  
in terms of eigenmodes on the CH space
\\ 
2. expansion coefficients of eigenmodes on the CH space 
that are extended onto the whole pseudosphere 
in terms of eigenmodes on the pseudosphere.  
\\
\\
\centerline{\bf Acknowledgments}
\\
\indent

I would like to thank my advisor Professor Kenji Tomita for his many helpful
discussions and continuous encouragement. 
I would also like to thank Professor Toshiro Matsumoto 
for his extensive advice on the 
boundary element methods and Professor Jeff Weeks and the Geometry
Center in University of Minnesota for providing me the data of CH
spaces. Numerical computation in this work was carried out by HP
Exemplar at the 
Yukawa Institute Computer Facility. 
 I am supported by JSPS Research Fellowships 
for Young Scientists, and this work is supported partially by 
Grant-in-Aid for Scientific Research Fund (No.9809834). 

\pagebreak
\appendix
\section{Boundary integral equation} 

\begin{figure}
\includegraphics[height=10cm,clip]{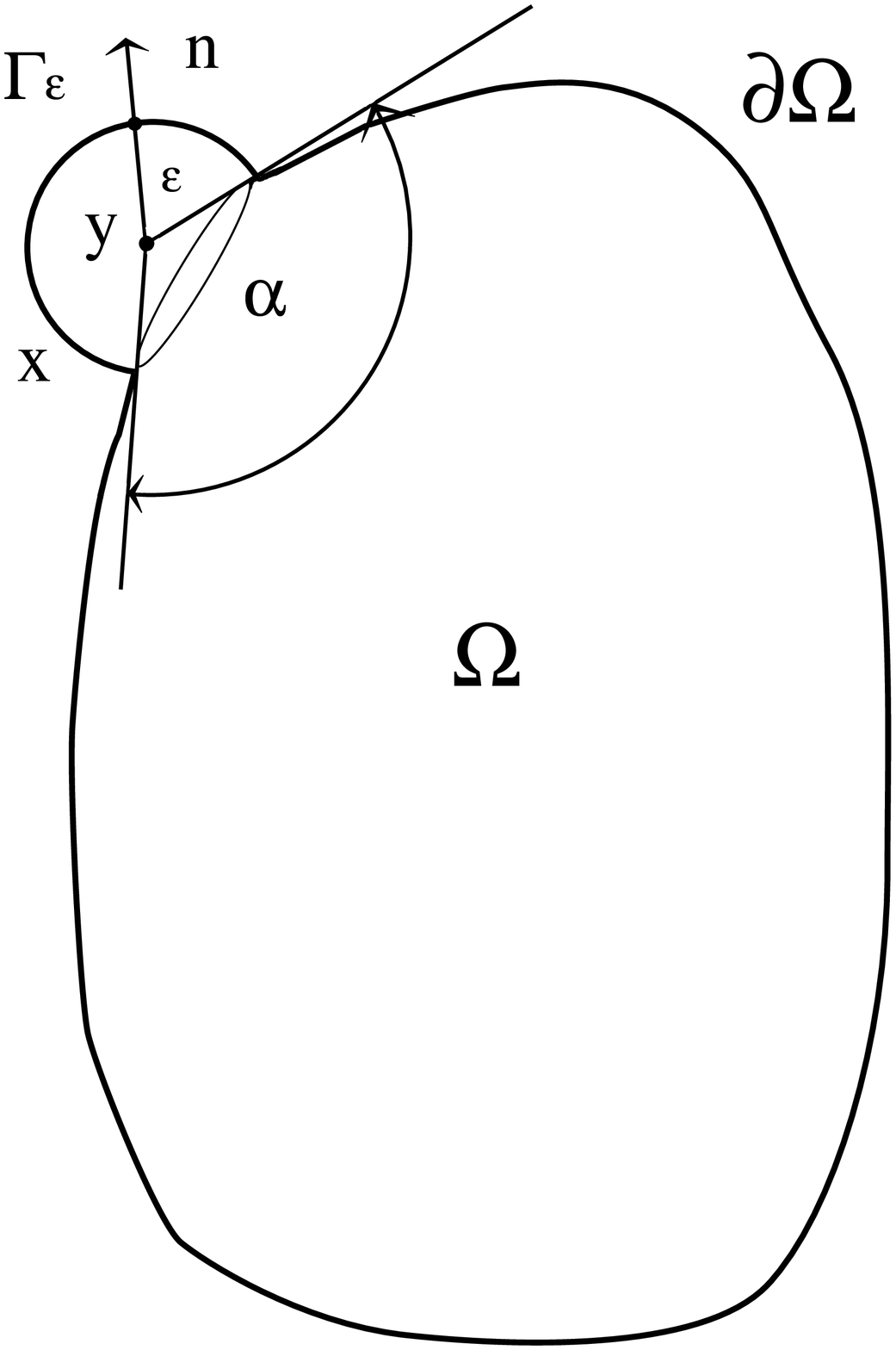}
\caption{Boundary integral}
\label{fig:BEMlimit}
\end{figure}

\indent
\indent
Here, we derive the boundary integral equation (\ref{eq:bem0}) in
section 1. For simplicity, we prove the formula in 3-spaces. 
\\
\indent
First, we start with Eq.(\ref{eq:inter}) with dimesnsion $M=3$.
Although the integrand in Eq.(\ref{eq:inter}) is divergent at 
$\x=\y\in \del\Omega$, the integration can be regularized as follows. 
Let us draw a sphere with center $\y\in \del\Omega$ with
small radius $\epsilon$ and let $\Gamma_\epsilon$ be the outer 
spherical boundary and $\alpha$ and $\beta$ be the internal solid angle 
and external solid angle as shown in figure \ref{fig:BEMlimit},
\begin{equation}
u(\y)
+\int_{\del\Omega+\Gamma_\epsilon} G_E(\x,\y) \nabla_i u \,\sqrt{g}\, dS^i
-\int_{\del\Omega+\Gamma_\epsilon} (\nabla_i G_E(\x,\y)) u \,\sqrt{g}\,
dS^i=0.
\label{eq:first}
\end{equation}
The singular terms in Eq.(\ref{eq:first}) can be separated 
from non-singular terms as
\begin{eqnarray}
\lim_{\epsilon \rightarrow 0}\int_{\del \Gamma+\Gamma_\epsilon}
G_E(\x,\y) \nabla_i u \,\sqrt{g}\, dS^i&=&\int_{\del \Gamma}
G_E(\x,\y) \nabla_i u \,\sqrt{g}\, dS^i
\N
\\
\N
&+&\lim_{\epsilon \rightarrow 0}
\int_{\Gamma_\epsilon}
G_E(\x,\y) \nabla_i u \,\sqrt{g}\, dS^i,
\\
\N
\lim_{\epsilon \rightarrow 0}\int_{\del \Gamma+\Gamma_\epsilon}
(\nabla_i G_E(\x,\y)) u \,\sqrt{g}\, dS^i&=&\int_{\del \Gamma}
(\nabla_i G_E(\x,\y)) u \,\sqrt{g}\, dS^i
\N
\\
\N
&+&\lim_{\epsilon \rightarrow 0}
\int_{\Gamma_\epsilon}
(\nabla_i G_E(\x,\y)) u \,\sqrt{g}\, dS^i.
\label{eq:separate}
\\
\end{eqnarray}
If $\epsilon$ is sufficiently small, the region enclosed by 
$\Gamma_\epsilon$ can be approximated as an Euclidean subspace. 
In this region, the asymptotic form of the free Green's function 
$G_E$ takes the form
\begin{equation}
\lim_{\x \rightarrow \y} G_E(\x,\y)= - \f{\exp(ikd)}{4 \pi d}=-\f{1}{4
\pi d}-\f{ik}{4 \pi}+{\cal{O}}(d),
\end{equation}
where $d$ is the Euclidean distance between $\x$ and $\y$.
Taking the spherical coordinates $(\epsilon,\theta,\phi)$ with center
$\y$, the singular terms in Eq.(\ref{eq:separate}) are estimated as 
\begin{eqnarray}
\lim_{\epsilon \rightarrow 0}
\int_{\Gamma_\epsilon}
G_E(\x,\y) \nabla_i u(\x) \,\sqrt{g}\, dS^i&=&
\lim_{\epsilon \rightarrow 0} - \int_{\beta} \f{1}{4 \pi \epsilon}
\f{\del u(\x)}{\del n} \epsilon^2 d \Omega=0,
\N
\\
\lim_{\epsilon \rightarrow 0}
\int_{\Gamma_\epsilon}
(\nabla_i G_E(\x,\y)) u(\x) \,\sqrt{g}\, dS^i&=&
\lim_{\epsilon \rightarrow 0} \int_{\beta} \f{1}{4 \pi \epsilon^2}
\f{\del \epsilon(\x)}{\del n} u(\x) \epsilon^2 d \Omega
\N
\\
&=&\f{\beta}{4\pi} u(\x),
\label{eq:limit}
\end{eqnarray}
where $d \Omega$ denotes the infinitesimal solid angle element. 
Taking the limit $\epsilon \rightarrow 0$ in Eq.(\ref{eq:first}), 
we have the boundary integral equation in 3-spaces,
\begin{equation}
\f{1}{4 \pi} \alpha(\y) u(\y)
+\int_{\del\Omega} G_E(\x,\y) \nabla_i u \,\sqrt{g}\, dS^i
-\int_{\del\Omega} (\nabla_i G_E(\x,\y)) u \,\sqrt{g}\, dS^i=0,
\label{eq:bemAPENDIX}
\end{equation}
where $\alpha(\y)$ denotes the internal solid angle at $\y$. If the
boundary is smooth at $\y$, $\alpha(\y)$ is equal to $2 \pi$ which
gives the coefficients $1/2$ in Eq.(\ref{eq:bem0}). Similarly, one can 
prove the formula for $M=2$ and $M>3$.
\pagebreak
\section{Three-dimensional hyperbolic spaces}
\indent

The discrete
subgroup $\Gamma$ of $PSL(2,{\M C})$ which is the orientation-preserving 
isometry group of the simply-connected hyperbolic 3-space ${\M H}^3$
is called the Kleinian group. Any CH space (either manifold or
orbifold) can be described 
as compact quotients ${\cal M}={\M
H}^3/ \Gamma$. The classification of the Kleinian group has not been
completed. However, several procedures for constructing compact
hyperbolic spaces are known. 
For further detail, see 
\cite{Thurston1,Thurston2,Beardon,Fomenko}.  
\\
\indent
The standard pseudospherical coordinates ($\chi, \theta,\phi$) for
${\M H}^3$ with curvature radius $R$ are given by
\begin{eqnarray}
X_0&=&R \cosh \chi,~~~X_1=R \sinh\chi \sin\theta \cos\phi
\nonumber
\\
X_2&=&R \sinh\chi \sin\theta \sin\phi, ~~~X_3=R \sinh \chi \cos \theta 
\end{eqnarray}
with
\begin{equation}
0 \leq\chi < \infty,~~0 \leq\theta < \pi,~~0 \leq\phi < 2\pi.
\end{equation}
In these coordinates, the line element takes the form
\begin{equation}
ds^2=\sum_{i=0}^3 (d X_i)^2=
R^2 \bigl[d \chi^2+ \sinh \chi^2(d \theta^2+\sin^2 \theta d
\phi^2)\bigr].
\end{equation}
\\
\indent
The Poincar$\acute{\textrm{e}}$ representation is
obtained by the transformation
\begin{equation}
x_1=R \tanh \f{\chi}{2} \sin \theta \cos \phi, ~~
x_2=R \tanh \f{\chi}{2} \sin \theta \sin \phi, ~~
x_3=R \tanh \f{\chi}{2} \cos \theta,
\end{equation}
which maps ${\M H}^3$ into the open ball 
$\{(x_1,x_2,x_3)\in {\M E}^3\,|\,x_1^2+x_2^2+x_3^2<R^2\}$ called 
the $Poincar\acute{\textrm{e}}~ ball$.  The line element in these coordinates 
takes the form 
\begin{equation}
ds^2=\f{4(dx_1^2+dx_2^2+dx_3^2)}{\biggl(1-\f{x_1^2+x_2^2+x_3^2}{R^2}
\biggr)^2},
\end{equation}
and the geodesic distance $d$ between $\x$ and $\x'$ is given by
\begin{equation}
\cosh \bigl[R^{-1}d(\x,\x') \bigr]= 1+\f{2 |\T{\x}-\T{\x}'|^2}{(1-|\T{\x}|^2)
(1-|\T{\x}'|^2)}.
\end{equation}
where $|~ |$ denotes the Euclidean norm and 
$\T{\x}=R^{-1}\x, ~~\T{\x}'=R^{-1}\x'$. Note that geodesics 
in the Poincar$\acute{\textrm{e}}$ ball are either diameters or
semi-circles which orthogonally intersect with the 
boundary of the Poincar$\acute{\textrm{e}}$ ball. 
\\
\indent In Poincar$\acute{\textrm{e}}$ coordinates, the metric is
conformally flat so that the computation of the boundary integral
equation becomes simpler.
\\
\indent
Another commonly used set of coordinates is obtained from the upper-half space
representation which is defined by the transformation 
\begin{eqnarray}
y_1&=&\f{\sinh \chi \sin \theta \cos\phi}{D},
y_2=\f{\sinh \chi \sin \theta \sin\phi}{D},
y_3=\f{1}{D},
\nonumber
\\
D&=&{\cosh\chi-\sinh\chi\cos\theta},
\end{eqnarray}
which maps ${\M H}^3$ into the upper-half space 
${\M E}^3_+ =\{(y_1,y_2,y_3)\in {\M E}^3\,|\,y_3>0\}$.
In these coordinates, the line element takes the form
\begin{equation}
ds^2=\f{R^2(dy_1^2+dy_2^2+dy_3^2)}{y_3^2}.
\end{equation}  
The geodesic distance is given by
\begin{equation}
\cosh \bigl[R^{-1}d(\y,\y')\bigr]= 1+\f{|\y-\y'|^2}{2 y_3 y'_3}.
\end{equation}
In the upper-half space model, geodesics are either straight 
vertical lines or semi-circles orthogonal to the boundary of 
the upper-half space. In this coordinates the metric is conformally
flat as in Poincar$\acute{\textrm{e}}$ coordinates.
If we represent a point $p$ on the upper-half space, as a quaternion whose
fourth component equals zero \cite{Beardon}, then the actions of  $PSL(2,{\M
C})$ on ${\M H}^3 \cup {\M C} \cup\{ \infty\} $ can be 
described by simple formulas, 
\begin{equation}
\T\gamma: p\rightarrow p'=\f{a p+ b}{c p+ d}, 
~~~~~ad-bc=1,~~~p\equiv z + y_3 {\bj}, ~~~z=y_1+y_2 {\bi},
\end{equation}
where a, b, c and d are complex numbers and $1$, $\bi$ and $\bj$ are
represented by matrices as,
\BE
1=
\left(
\begin{array}{@{\,}cc@{\,}}
1 & 0\\ 
0 & 1 
\end{array}
\right),~~~
\bi=
\left(
\begin{array}{@{\,}cc@{\,}}
i & 0\\ 
0 & -i 
\end{array}
\right),~~~
\bj=
\left(
\begin{array}{@{\,}cc@{\,}}
0 & 1\\ 
-1& 0 
\end{array}
\right).
\EE
As $\overline{p}=z-y_3 \bj$, the action $\T\gamma$ is
explicitly written as   
\begin{eqnarray}
\T\gamma:{\M H}^3\cup{\M C}\cup\{ \infty \} &\rightarrow& 
{\M H}^3\cup{\M C}\cup\{ \infty \},
\nonumber
\\
\nonumber
\\
\T\gamma:(z(y_1,y_2),y_3)~~ &\rightarrow& 
\Biggl
( \f{(az+b)(\overline{cz+d})+a\bar{c}y_3^2}{|cz+d|^2+|c|^2y_3^2},
\f{y_3}{|cz+d|^2+|c|^2y_3^2}\Biggr). 
\end{eqnarray}
If we restrict the action $\T\gamma$ on ${\M C}\cup\{ \infty \}$,
or equivalently, $y_3=0$, the action is described as
\begin{equation}
\gamma: z\rightarrow z'=\f{a z+ b}{c z+ d}, 
~~~~~ad-bc=1,~~~z\equiv y_1+i y_2,
\end{equation}
$\gamma$ is called the \ti{M}$\ddot{o}$\ti{bius} \ti{transformation}, 
and $\T \gamma$ is
called the \ti{extended} \ti{M}$\ddot{o}$\ti{bius} \ti{transformation}.
\\ 
\\
\indent
In the $Klein$ (projective) model, 
the geodesics and planes are mapped into
their Euclidean counterparts. Since the fundamental domain is
enclosed by Euclidean planes in the Klein coordinates, the task of 
generating meshes is much
easier than other coordinates.  
The transformation  
\begin{equation}
z_1=R \tanh \chi \sin \theta \cos \phi, ~~
z_2=R \tanh \chi \sin \theta \sin \phi, ~~
z_3=R \tanh \chi \cos \theta
\end{equation} 
can be understood as the projection of the
hyperboloid ($X_0,X_1,X_2,X_3$) onto the interior of the sphere 
($R,z_1,z_2,z_3$) along lines originating from the origin (0,0,0,0).  
The geodesic distance can be represented as 
\begin{equation}
\cosh \bigl[R^{-1}d(\z,\z')\bigr]= \f{1-\T{\z} \cdot \T{\z}'}
{\sqrt{(1-|\T{\z}|^2)(1-|\T{\z}'|^2)}}.
\end{equation}
where $\cdot$ denotes the Euclidean inner product 
and $\T{\z}=R^{-1}\z, ~~\T{\z}'=R^{-1}\z'$.
\\
\\
\indent
The possible values for the volume of the CH manifolds 
are bounded below and no upper bound exists.
The minimal value has not yet been known, although Gabai et al have
proved that 
$Vol_{\textrm{min}}>0.16668...R^3\cite{homotopy}$.  
Thurston proposed a manifold $Q_2$ as 
a candidate for
the three-dimensional hyperbolic manifold of the minimum volume 
$Vol(Q_2)=0.98139R^3$ \cite{SnapPea}. However, Weeks \cite{Weeks1} 
and independently, Matveev and Fomenko
\cite{Matveev} discovered a CH manifold $Q_1$ with the 
smallest value $Vol(Q_1)=0.94272R^3$ in the known CH manifolds 
and it is conjectured 
to be the one with the minimum volume. A computer program 
''SnapPea'' by Jeff Weeks \cite{Hodgson} has made it possible to catalog
and study a large number of CH and non-CH spaces which include
$Q_1$,$Q_2$ and thousands of cusped and non-cusped hyperbolic 
3-manifolds.
\\
\indent
\BF
\IG[width=12cm,clip]{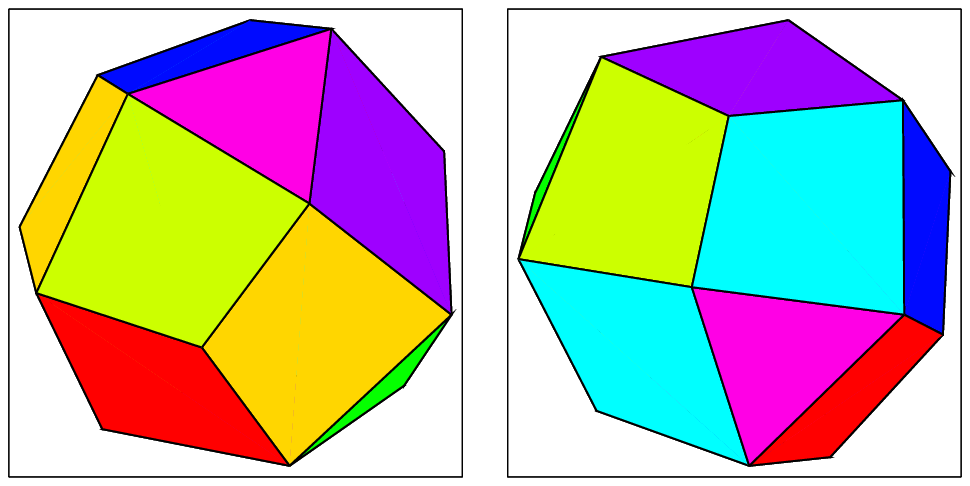}
\caption{Fundamental domain of the Thurston manifold
which is  
viewed from a point (0,0,-1000) and a point (0,0,1000) in the 
Klein coordinates, respectively.}
\label{fig:FD}
\EF
 Let us see how CH manifolds are
characterized in the SnapPea. Any element of the discrete isometry
group $\Gamma$ which is equivalent to the fundamental group $\pi_1({\cal 
M})$ can be described as a $word$ which is a product of generators
$\{g_1,\ldots,g_s\}$,  
\begin{equation}
~~~~~~~~~~~~~~~g=g^{n_{1}}_{m_{1}}\ldots g^{n_{j}}_{m_{j}},~~(j,n_{j}\in Z,~~m_{j}=1,~\ldots ,s),
~~~g \in \Gamma. 
\end{equation}
The above expression is not unique, since they are subject to a set of 
relations, each of which takes the form,
\begin{equation}
{\cal I}=g^{k_{1}}_{l_{1}}\ldots g^{k_{j}}_{l_{j}},~~~
(j,k_{j}\in Z,~~l_{j}=1,~\ldots ,s),
\end{equation}
where ${\cal I}$ denotes the identity.
Note that different expression of $g$ is possible by choosing
different generators.   
In the case of Thurston's manifold $Q_2$, 
$\Gamma$ has a simple presentation,     
\begin{equation}
\Gamma=\{a,b~:~ab^3aba^{-2}b,~~ab^{-1}a^{-1}baba^{-1}b^{-1}ab^{-1}\},
\end{equation}
where $a$ and $b$ are generators and words in the parenthesis
are equal to identities. 
This representation is simple for describing $\Gamma$ but not
convenient for describing the fundamental domain. 
Choosing a coordinate system centered at a point of locally maximum of
the injectivity radius\footnote{The injectivity radius of a point $p$ is 
equal to half the length of the shortest periodic geodesic on $p$.},
generators which define the face identification maps in the
pseudospherical coordinates can be described 
by 8 matrices (see appendix C),
which implies that the number of the faces on the boundary of the 
fundamental domain is sixteen. For instance, the center
$(X_0,X_1,X_2,X_3)=(1,0,0,0)$ is moved to a point 
$(1.63032,-0.5657,0.993147,-0.592943)$ by $T_1$.     
The fundamental domain of the Thurston manifold can be computed from
these 8 matrices. First, let us make 16 copies of the basepoint
$(1,0,0,0)$ that are obtained by multiplying the basepoint by 
8 matrices and their 
inverse matrices. Next, the basepoint and 
the 16 copies are connected with 16 geodesic segments. 
If one put planes on 
the equidistant points on the
segments at right angle, then one obtains the fundamental domain 
enclosed by the 16 planes.
The face 
identifications are shown in figure \ref{fig:FD}, in which 
each color of the faces
corresponds to one of the identification maps. For instance, a 
point on the face with red color in the left figure is identified with
the corresponding point on the face with the same color 
in the right figure by $T_1$.   
\pagebreak
\section{Table of matrices}
In the Minkowski coordinates
$(t,x,y,z)$, the 8 generators which define the
fundamental domain of the Thurston manifold with the basepoint
$(1,0,0,0)$ are 
described by the following 8 matrices,
\begin{eqnarray}
T_1=\pmatrix{ 1.630319018827 & 
  -0.823154893134 & 
  0.949658853916 & 
  -0.280185985083 \cr 
  -0.565700099811 & 
  0.926953540114 & 
  0.03235859132 & 
  0.678031458683 \cr 
  0.993147485366 & 
  -0.896311100897 & 
  1.03061185404 & 
  0.34757379707 \cr 
  -0.592943144435 & 
  -0.122341828367 & 
  -0.915775117974 & 
  0.705669882356 \cr  }, 
\N
\\
\N
\\
T_2=\pmatrix{ 1.630319018827 & 
  -0.404520784012 & 
  -0.353128575811 & 
  -1.170300494487 \cr 
  -0.725744142114 & 
  1.032680902406 & 
  0.430856766432 & 
  0.524058355949 \cr 
  -0.749088496222 & 
  0.153460453167 & 
  -0.490988659119 & 
  1.138645511607 \cr 
  0.755050970573 & 
  0.271398430022 & 
  -0.835459379318 & 
  -0.893561685534 \cr  },
\N
\\
\N
\\
T_3=\pmatrix{ 1.630319018827 & 
  0.687449077 & 
  -0.510375368061 & 
  -0.961702060597 \cr 
  -0.493618131693 & 
  0.144914737528 & 
  -0.255278318146 & 
  1.075867816726 \cr 
  0.044016981085 & 
  0.704808911001 & 
  0.708779194553 & 
  0.053046648969 \cr 
  -1.188420695119 & 
  -0.977154210597 & 
  0.832435012465 & 
  0.874394274525 \cr  },
\N
\\
\N
\\
T_4=\pmatrix{ 1.630319018827 & 
  -0.076636501663 & 
  -1.186821280266 & 
  -0.493479683947 \cr 
  0.990384655015 & 
  0.188025082259 & 
  -1.393877403859 & 
  0.051128429897 \cr 
  -0.430529153566 & 
  0.973085065179 & 
  0.444575008386 & 
  0.202024429006 \cr 
  -0.701229624437 & 
  -0.153704839725 & 
  0.517690637565 & 
  1.09548811596 \cr  },
\N
\\
\N
\\
T_5=\pmatrix{ 1.630319018827 & 
  -0.422808564099 & 
  -0.669242281726 & 
  1.015523406733 \cr 
  -1.169223329346 & 
  0.245239342961 & 
  1.092969619938 & 
  -1.054683966146 \cr 
  -0.48230025187 & 
  1.00284583704 & 
  -0.146468308544 & 
  -0.453277833869 \cr 
  -0.241336645178 & 
  -0.336043112815 & 
  -0.481507711397 & 
  -0.84467077995 \cr  },
\N
\\
\N
\\
T_6=\pmatrix{ 1.630319018827 & 
  -1.258320677027 & 
  -0.085154694975 & 
  -0.259456845811 \cr 
  -0.334091294985 & 
  0.339168543325 & 
  0.989853092692 & 
  0.129508870322 \cr 
  -1.230309911716 & 
  1.513335320951 & 
  -0.161421439076 & 
  0.444321735047 \cr 
  -0.180722524595 & 
  0.422080361796 & 
  0.03721956436 & 
  -0.92364684978 \cr  },
\N
\\
\N
\\
T_7=\pmatrix{ 1.815210205228 & 
  -1.011761093044 & 
  -1.117355434627 & 
  0.15114368155 \cr 
  -1.394305392881 & 
  1.36137877709 & 
  0.985732513048 & 
  -0.345060526033 \cr 
  0.587810577053 & 
  0.152666848297 & 
  -1.125213789716 & 
  -0.236871347696 \cr 
  0.073343616291 & 
  -0.383407313596 & 
  0.103481918443 & 
  -0.920689747139 \cr  },
\N
\\
\N
\\
T_8=\pmatrix{ 1.815210205228 & 
  -0.288909259155 & 
  0.774531377414 & 
  -1.269496228642 \cr 
  0.093635040681 & 
  0.602143185543 & 
  -0.536756936203 & 
  -0.598400448183 \cr 
  -0.930634634266 & 
  0.833203272177 & 
  -0.104192832263 & 
  1.077495699961 \cr 
  -1.191696163383 & 
  -0.163292533092 & 
  -1.1405852444 & 
  1.045246666925 \cr  }.
\N
\end{eqnarray}  
\end{document}